\documentclass[times,authoryear]{elsarticle}

\usepackage{jasr}
\usepackage{framed,multirow}

\usepackage{amssymb}
\usepackage{latexsym}

\usepackage[switch]{lineno}

\usepackage{url}
\usepackage{xcolor}
\usepackage{tabularx}
\definecolor{newcolor}{rgb}{.8,.349,.1}
\usepackage{footmisc}

\usepackage[citebordercolor=white]{hyperref}

\journal{Advances in Space Research}

\begin{document}

\verso{Nandi \textit{et.al.}}

\begin{frontmatter}

\title{Shaping SHAPE - A spectro-polarimeter onboard Chandrayaan-3 to observe Earth as an Exoplanet}%

\author[1]{Anuj \snm{Nandi}\corref{cor1}}
\cortext[cor1]{ 
  Email: anuj@ursc.gov.in}
\author[1]{Swapnil \snm{Singh}\corref{cor2}}
\cortext[cor2]{ 
  Email: swapnils@ursc.gov.in}
\author[1]{Bhavesh \snm{Jaiswal}}
\author[2]{Anand \snm{Jain}}
\author[2]{Smrati \snm{Verma}}
\author[1]{Reenu \snm{Palawat}}
\author[1]{Ravishankar \snm{B.T.}}
\author[1]{Brajpal \snm{Singh}}
\author[2]{Priyanka \snm{Das}}
\author[2]{Supratik \snm{Bose}}
\author[2]{Supriya \snm{Verma}}
\author[2]{Waghmare Rahul \snm{Gautam}}
\author[2]{Yogesh Prasad \snm{K. R.}}
\author[3]{Bijoy \snm{Raha}}
\author[3]{Bhavesh \snm{Mendhekar}}
\author[3]{Sathyanaryana \snm{Raju K.}}
\author[3]{Srinivasa Rao \snm{Kondapi V.}}
\author[2]{Sumit \snm{Kumar}}
\author[2]{Mukund Kumar \snm{Thakur}}
\author[1]{Vinti \snm{Bhatia}}
\author[1]{Nidhi \snm{Sharma}}
\author[1]{Govinda Rao \snm{Yenni}}
\author[1]{Neeraj Kumar \snm{Satya}}
\author[1]{Venkata \snm{Raghavendra}}
\author[2]{Vivechana \snm{M. S.}}
\author[2]{Evangelin Leeja \snm{Justin}}
\author[2]{Praloy \snm{Karmakar}}
\author[2]{Naga \snm{Manjusha J.}}
\author[2]{Motamarri \snm{Srikanth}}
\author[2]{Abhishek \snm{Singh}}
\author[2]{Prabakaran \snm{B.}}
\author[4]{Honey \snm{Gupta}}
\author[1]{Priyanka \snm{Mishra}}
\author[2]{Chinmay Kumar \snm{Rajhans}}
\author[2]{Kalpana \snm{K.}}
\author[5]{Veeramuthuvel \snm{P.}}

\affiliation[1]{organization={ISRO Satellite Integration and Test Establishment (ISITE), U. R. Rao Satellite Centre (URSC)},
                addressline={Indian Space Research Organisation (ISRO)},
                city={Bengaluru},
                postcode={560037},
                country={India}}
\affiliation[2]{organization={U. R. Rao Satellite Centre (URSC)},
                addressline={Indian Space Research Organisation (ISRO)},
                city={Bengaluru},
                postcode={560017},
                country={India}}
\affiliation[3]{organization={Laboratory for Electro-Optics Systems (LEOS), U. R. Rao Satellite Centre (URSC)},
                addressline={Indian Space Research Organisation (ISRO)},
                city={Bengaluru},
                postcode={560058},
                country={India}}                                       
\affiliation[4]{organization={ISRO Telemetry Tracking and Command Network (ISTRAC)},
                addressline={Indian Space Research Organisation (ISRO)},
                city={Bengaluru},
                postcode={560058},
                country={India}}
\affiliation[5]{organization={ISRO Headquarters},
                addressline={Indian Space Research Organisation (ISRO)},
                city={Bengaluru},
                postcode={560094},
                country={India}}   

\received{DD Month YYYY}
\finalform{DD Month YYYY}
\accepted{DD Month YYYY}
\availableonline{DD Month YYYY}
\communicated{S. Singh}

\begin{abstract}
\textbf{S}pectro-polarimetry of \textbf{HA}bitable \textbf{P}lanet \textbf{E}arth (SHAPE) is an experimental instrument on-board the Propulsion Module (Orbiter) of  Chandrayaan-3 mission. SHAPE is configured and designed to carry out disc-integrated \lq spectro-polarimetric' observations of Earth from the Lunar Orbit (LO) as well as from Highly Elliptical Orbits (HEO) around Earth. The instrument, which consists of three subsystems - Electro-Optical Detector System (EODS)-Optics, EODS-Electronics and Radio Frequency Source (RFS), is a compact and light-weight spectro-polarimeter with an Acousto-Optic Tunable Filter (AOTF) as the filtering element. The AOTF is driven by a radio frequency (80 $-$ 135 MHz) signal generated with the in-house developed RFS system. On application of the RF signal, the AOTF filters incident light in the Near-Infrared (NIR) band ($1.0 - 1.7$ \textmu m) and produces two diffracted narrow-band beams, which are linearly polarized in mutually perpendicular directions. The instrument optics with an FOV of $\sim2.6^{\mathrm{o}}$ is configured to focus two output beams onto two Indium-Gallium-Arsenide (InGaAs) detectors. A spectral resolution of $2 - 4$ nm has been achieved in the operating wavelength range with an in-house designed low-noise front-end electronics. The instrument is also equipped with processing and power electronics to process the signal, drive the electronics and bias the detectors with required voltages. In this work, we present the overall design aspects, results from the pre-launch ground-based tests, and in-orbit operations. With the present design configuration, the instrument has the capability to measure disc-integrated signatures of Earth for a range of phase angles, serving as a test bed to benchmark future observations of Earth-like exoplanets.
\end{abstract}

\begin{keyword}
\KWD Earth\sep Exoplanet\sep Spectro-polarimetry\sep Near-Infrared
\end{keyword}

\end{frontmatter}


\section{Introduction}
\label{sect:intro}  

The Chandrayaan-3 mission \citep{Ch3paper} was designed and configured with two main components: a Propulsion Module (PM) and a Lander Module (LM), with the primary mission objective of achieving a safe and soft landing on the Moon. The PM's main role was to carry the LM to a Lunar Orbit (LO) and the LM is integrated with a Rover intended for surface exploration. The LM and Rover were equipped with six payloads to conduct in-situ scientific experiments on the Moon's surface. The PM also carried a scientific payload named \textbf{S}pectro-polarimetry of \textbf{HA}bitable \textbf{P}lanet \textbf{E}arth (SHAPE), which was intended to operate after the LM's separation in order to observe Earth from lunar orbit. The mission was launched on 14 July 2023. Chandrayaan-3 entered lunar orbit on 5 August 2023, the LM separated from the PM on 17 August 2023 and subsequently performed a successful soft landing on the lunar surface on 23 August 2023\footnotemark[1] post which the lander payloads were operated (\citealp{Ch3lander, Ch3apxs}).
\footnotetext[1]{\url{https://www.isro.gov.in/Chandrayaan3.html}}

The SHAPE experiment aims to observe Earth as an exoplanet, i.e., through disc-integrated observations \citep{2022JATIS...8d4007J}. Since exoplanets cannot typically be spatially resolved by current observational facilities, their detected signals represent the integrated contribution from the entire planetary disc. To simulate such observations, SHAPE is designed to acquire spatially unresolved measurements of Earth from a distant vantage point. A lunar orbit is an ideal location for this purpose, as the Earth appears relatively small from that distance and subtends an angular diameter of approximately $\sim 2^{\mathrm{o}}$. The instrument Field of View (FOV) was therefore selected to closely match the apparent size of the Earth while minimizing the inclusion of surrounding background space. This configuration enables the collection of spectro-polarimeteric signal from the entire Earth disc, analogous to the integrated signal that would be obtained from a distant exoplanet to generate benchmark datasets for interpreting future exoplanet observations. For this reason, the SHAPE payload was accommodated on the PM. Figure 1 shows the PM of the Chandrayaan-3 spacecraft with the SHAPE payload mounted on the $-$PITCH panel and oriented along the +ROLL axis as the viewing direction. On the right side of the figure, a zoomed in 3D model of the SHAPE payload is shown with its three main components labelled. The payload was first operated on 20 August 2023.

\begin{figure}[h!]
\centering
\includegraphics[scale=0.33]{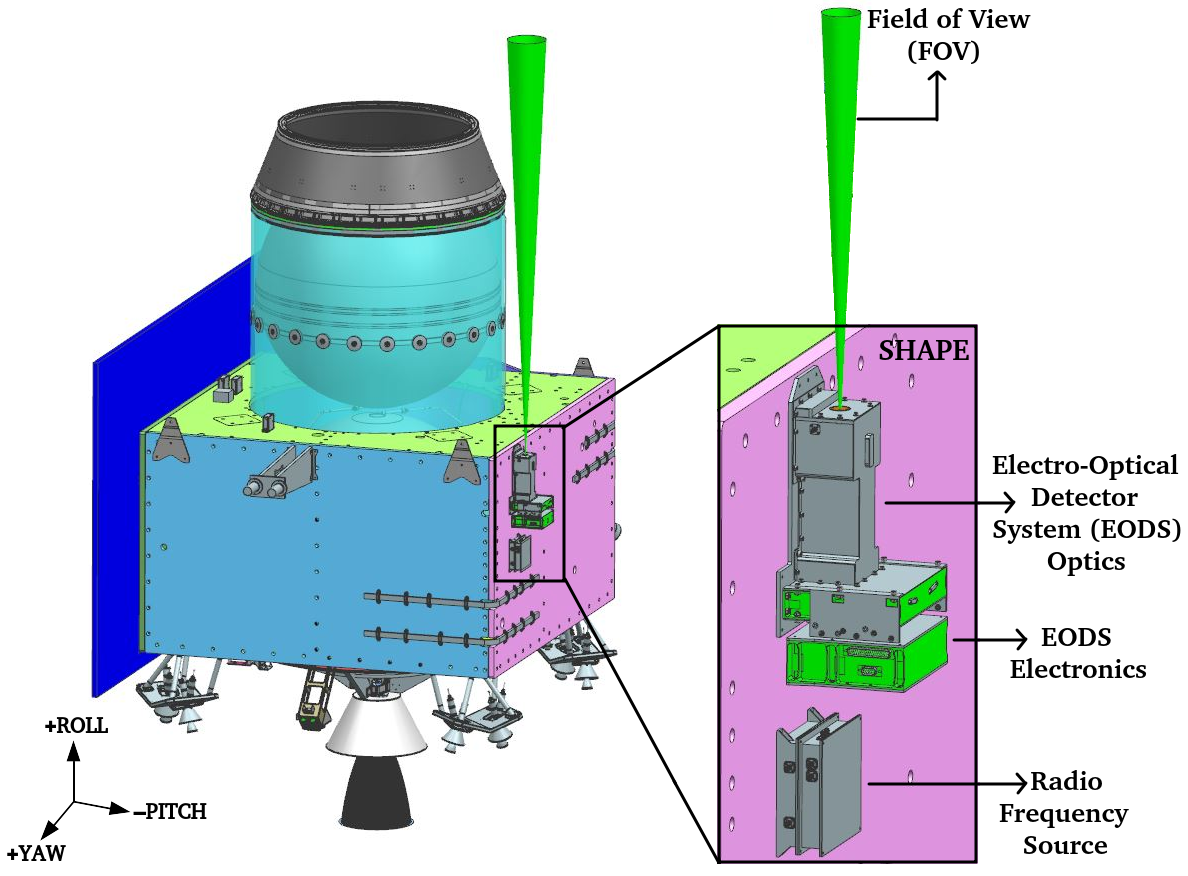}
\caption{The PM showing the mounted SHAPE instrument (left) along with the zoomed in view (right). Three major sub-systems are also marked.}
\label{fig:ch3pm}
\end{figure}

Observing Earth from a large distance, such as the Lunar orbit, provides a unique and scientifically valuable opportunity to study Earth as an unresolved, disc-integrated object, similar to how exoplanets are observed. Such orbits enable full disc measurements that capture Earth’s full diurnal cycle, phase angle variations, and cloud dynamics, which cannot be achieved from Low Earth Orbit (LEO) or geostationary platforms. Existing Earth observation missions, such as Moderate Resolution Imaging Spectroradiometer (MODIS; providing broadband spectroscopic data; \citealp{1993modis}) and POLarization and Directionality of the Earth's Reflectances (POLDER; offering limited polarimetric measurements; \citealp{1994ITGRS..32..598D}) sensor onboard the PARASOL satellite, which operated in a polar sun-synchronous orbit and provided multi-angle, polarized measurements of Earth, are constrained by their vantage points and observational strategies, which yield high-resolution local data but do not capture the global, instantaneous view. Attempts to reconstruct global properties by stitching together local observations are limited by phase angle variability and rapid atmospheric changes. Previous observations of Earth from deep space by interplanetary missions such as Voyager, Galileo, and EPOXI have demonstrated the feasibility of Earth-as-an-exoplanet measurements, but these efforts were limited to a few spectral bands and brief time intervals \citep{1977SSRv...21...75S, 1993Natur.365..715S,2011AsBio..11..907L}. A broadband spectro-polarimeter positioned in a LO would overcome these limitations by enabling continuous, phase-resolved, and disc-integrated observations of Earth. Such measurements are crucial both for advancing Earth atmospheric science and for validating models used in the characterization of Earth-like exoplanets \citep{2012Natur.483...64S, 2018AsBio..18..739F}. There are several proposed missions, such as the Lunar Observatory for Unresolved Polarimetry of Earth (LOUPE; \citealp{2021RSPTA.37990577K}), which aim to carry out such studies.

To address these gaps in exoplanet science, the SHAPE payload onboard Chandrayaan-3 is a significant step forward. Operating from the LO and from Highly Elliptical Orbit (HEO) around Earth, SHAPE is uniquely positioned to obtain spectro-polarimetric measurements of Earth as an unresolved, disc-integrated source. In contrast to previous opportunistic observations of Earth from interplanetary flybys, which were constrained by limited spectral coverage and short temporal baselines, SHAPE provides long-term monitoring of Earth’s diurnal cycle and phase-angle variations in the NIR band. These measurements serve as critical benchmarks for the validation of retrieval techniques and forward models employed in the characterization of Earth-like exoplanets. SHAPE employs an Acousto-Optic Tunable Filter (AOTF) to carry out the spectro-polarimetric measurements. AOTFs have been widely utilized in space-based instruments for the study of planetary atmospheres due to their rapid tunability and absence of moving mechanical components (\citealp{2006JGRE..111.9S03K, 2012P&SS...65...38K, 2017SPIE10562E..1MK}). Preliminary details of the payload, along with one of the first-light spectra, are presented in brief by \cite{2024arXiv241207416N}. In the present work, we provide a comprehensive account of the instrument's design, ground testing, and in-orbit operations.

The manuscript is subsequently organized into 6 major sections. Section 2 presents a scientific overview of the instrument, outlining the primary objectives and the motivation behind its development. Section 3 details the overall design configuration, with a detailed description of the three sub-systems and their design. Section 4 describes the mechanical and thermal design aspects, including structural configuration and thermal control strategies implemented to ensure in-orbit stability. Section 5 discusses all the ground-based characterization, qualification and calibration tests carried out to validate the instrument under simulated launch and space conditions. Section 6 outlines the in-flight operation strategy and initial observations, highlighting the instrument's performance in orbit. Finally, Section 7 provides a summary of the major findings and concludes the paper.

\section{Scientific Overview} 
Earth serves as the only available ground-truth benchmark providing a unique opportunity to validate exoplanet characterization techniques against a planet whose atmospheric and surface properties are already known. Observing Earth as an exoplanet has been a longstanding goal in planetary science, aiming to understand how exoplanets will appear to observers using various techniques and instruments, both on-ground and in-space. These disk-integrated spectro-polarimetric observations of Earth capture effects such as phase-dependent variability and polarization signatures that cannot be fully reproduced by simply degrading existing high-resolution Earth observations. Early attempts, such as those by the Galileo, Mars Global Surveyor and Voyager missions (\citealp{1993Natur.365..715S, 1997JGR...10210875C}), utilized limited observations from interplanetary distances, providing only a snapshot of Earth's appearance without detailed spectral or polarimetric data. Most Earth observation satellites operate in Sun-synchronous low-Earth orbits, limiting observations to a portion of the Earth's disk as well as specific viewing angles and times of day. POLDER provided multi-angle, multi-spectral polarisation measurements of Earth’s atmosphere and surface, helping to characterize aerosols, clouds, and surface reflectance. Its data were instrumental in detecting features such as Rayleigh scattering, cloud cover, and surface signatures, including vegetation and ocean colour \citep{1994ITGRS..32..598D}. EPOXI observations of distant Earth offer time-resolved, multi-wavelength visible photometry and NIR spectroscopy of Earth's disk over several days, capturing a complete 24-hour rotation period \citep{2011AsBio..11..907L}. Apart from in-orbit observations, Earthshine observations have provided more detailed insights into Earth's characteristics as an exoplanet (\citealp{2002ApJ...574..430W, 2004AdSpR..34..293M, 2012Natur.483...64S, 2014A&A...562L...5M, 2019A&A...622A..41S, 2021A&A...653A..99T}). Earthshine is the light reflected from Earth's atmosphere and surface, which illuminates the Moon's dark side and is then reflected back to Earth. These studies have employed high-resolution spectro-polarimetry to obtain polarisation spectra across various phase angles, revealing spectral and temporal variability in Earth's polarisation. But these observations are limited by their faintness, dependence on lunar phase, atmospheric interference, etc. These studies highlight the importance of spectro-polarimetry in characterizing Earth as an exoplanet.

SHAPE carries out ‘first of its kind’ disc-integrated spectro-polarimetric observations of Earth. The payload has the capability to observe Earth's disc-integrated polarimetric signatures across a wide spectral range at various phase angles. SHAPE data will unlock the answers about the disc-integrated spectrum and polarisation of an Earth-like exoplanet and also how these features vary with the planet's rotation and revolution. To accomplish this, data from SHAPE will be used to carry out the following studies.

\subsection{Spectroscopic Studies} 
The presence of features of various gases, land and ocean can give us clues about the conditions on that planet. SHAPE studies the spectral fingerprints of various biosignatures (mainly H$_2$O, O$_2$ and CO$_2$) and cloud reflectance in the NIR wavelength range. Earthshine observations indicate that the reflection spectrum in the $0.7-2.4$ \textmu m range is dominated by signatures of H$_2$O, O$_2$, CO$_2$, and clouds \citep{2006ApJ...644..551T}. The O$_2$ 1.27 \textmu m band is relatively weak compared to the stronger optical O$_2$ A-band and primarily probes the middle and upper atmosphere. Owing to its spectral proximity to CO$_2$ band (between 1.57 \textmu m and 1.61 \textmu m), it provides a valuable proxy for atmospheric path length and serves as a diagnostic of atmospheric structure, dynamics, and column density \citep{2023Bai}. The clouds, though spectrally flat in the visible, are distinguishable in the NIR and critical for differentiating between ice and water clouds \citep{1998JApMe..37.1421G}, since the upper level clouds are made of ice and the lower and middle level clouds are made of water. In particular, the NIR H$_2$O band, centred near $\sim$1.4 \textmu m, exhibits a broad, structured absorption arising from a dense forest of rotational$-$vibrational transitions. These bands are modulated by cloud optical depth and cloud-top altitude as shown by \cite{2006ApJ...644..551T}. The temporal variations in the H$_2$O band depths can also be used to study how the local changes (like monsoon) manifest in the disc-integrated spectrum and whether these changes are strong enough to comment on the variations in different cloud layers. Spectroscopic results from SHAPE will be directly applied to study the cloud cover for characterizing a habitable exoplanet where the presence of water is expected.

\subsection{Polarimetric Studies} 
Polarisation signatures from Earth seen in the NIR band are due to the scattering from the clouds, with some contribution from sharp reflections from reflecting surfaces. The disc-integrated polarisation will have components from various parts of Earth. SHAPE observations will be used to study the contribution of each of these features of the planet. SHAPE measures linear polarization because it is the dominant polarization signature in reflected light from Earth-like planets, arising primarily from atmospheric Rayleigh and Mie scattering \citep{1974SSRv...16..527H}. Since linear polarization is significantly stronger than circular polarization and carries key information about atmospheric and cloud properties, it is the most sensitive observable for planetary characterization. It will further help in segregating the clouds based on altitude, as the clouds made of ice would have different polarisation signatures than the clouds which are made of liquid water (due to the difference in the refractive index and the shape of cloud droplets). \cite{2021A&A...653A..99T} conducted NIR polarimetric Earthshine observations, identifying a positive correlation between polarisation degree and ocean fraction, suggesting that polarimetric signatures can be indicative of Earth's surface properties. Polarisation measurements are highly sensitive to specular reflections from smooth surfaces and can thus be effectively used to study the phenomenon of glint $-$ the direct reflection of sunlight from ocean surfaces. In disc-integrated observations, the detection of glint serves as a promising diagnostic tool for inferring the presence of surface liquid water, particularly oceans, on Earth-like exoplanets (\citealp{2008Icar..195..927W, 2010ApJ...721L..67R}). The concept has been explored using limited observations from instruments such as POLDER (\citealp{1994ITGRS..32..598D, 1998GeoRL..25.1879B}). However, these observations were restricted in scope due to orbital constraints and instrument capabilities, underscoring the need for dedicated platforms capable of monitoring glint across a broader range of viewing geometries and phase angles. SHAPE is also designed to detect polarisation signatures associated with oceanic glint and to characterize cloud properties through their distinct polarimetric signatures. SHAPE conducts detailed measurements of the degree of linear polarisation within the H$_2$O absorption band and its dependence on planetary phase angle. This is extremely useful for future disc-integrated exoplanet observations, where polarisation signatures can constrain atmospheric scattering and cloud micro-physical properties.

\begin{table}[h!]
\begin{center}
\caption{Summary of science design requirements and instrument capabilities.}
\label{table:scireq}
\renewcommand{\arraystretch}{1.3} 
\begin{tabularx}{\textwidth}{|>{\centering\arraybackslash}c|>{\centering\arraybackslash}X|>{\centering\arraybackslash}X|}
\hline
\textbf{Science} & \textbf{Desired Requirements} & \textbf{SHAPE Capabilities} \\ \hline
\multirow{4}{*}{Gaseous species (H$_2$O, O$_2$, CO$_2$)} 
& Spectral range of 1.0$-$1.7~\textmu m 
& Spectral range of 1.04$-$1.70~\textmu m \\ \cline{2-3}

& Detector + Electronics Noise $<$ 2.6 mV 
& Detector + Electronics Noise $<$ 2.6 mV \\ \cline{2-3}

& SNR $>$100 with variable integration times (10 to 1000 ms) 
& SNR $>$100 with variable integration times (10 to 1000 ms) \\ \cline{2-3}

& Multiple scan modes: Full scan and Window scan 
& Multiple scan modes: Full scan and Window scan \\ \hline

\multirow{3}{*}{Cloud properties, Glint} 
& Polarimetric accuracy $\leq$ 1\% 
& Polarimetric accuracy $<$ 1\% (Band or $\triangle$ polarisation) \\ \cline{2-3}

& Two perpendicularly polarized chains to estimate linear polarisation 
& AOTF output is two mutually perpendicularly polarized beams \\ \cline{2-3}

& SNR $>$100 with variable integration times (10 to 1000 ms) 
& SNR $>$100 with variable integration times (10 to 1000 ms) \\ \hline

\end{tabularx}
\end{center}
\end{table}

\vspace{-5mm}

\section{Design Configuration}
\begin{figure}[h!]
\centering
\includegraphics[scale=0.5]{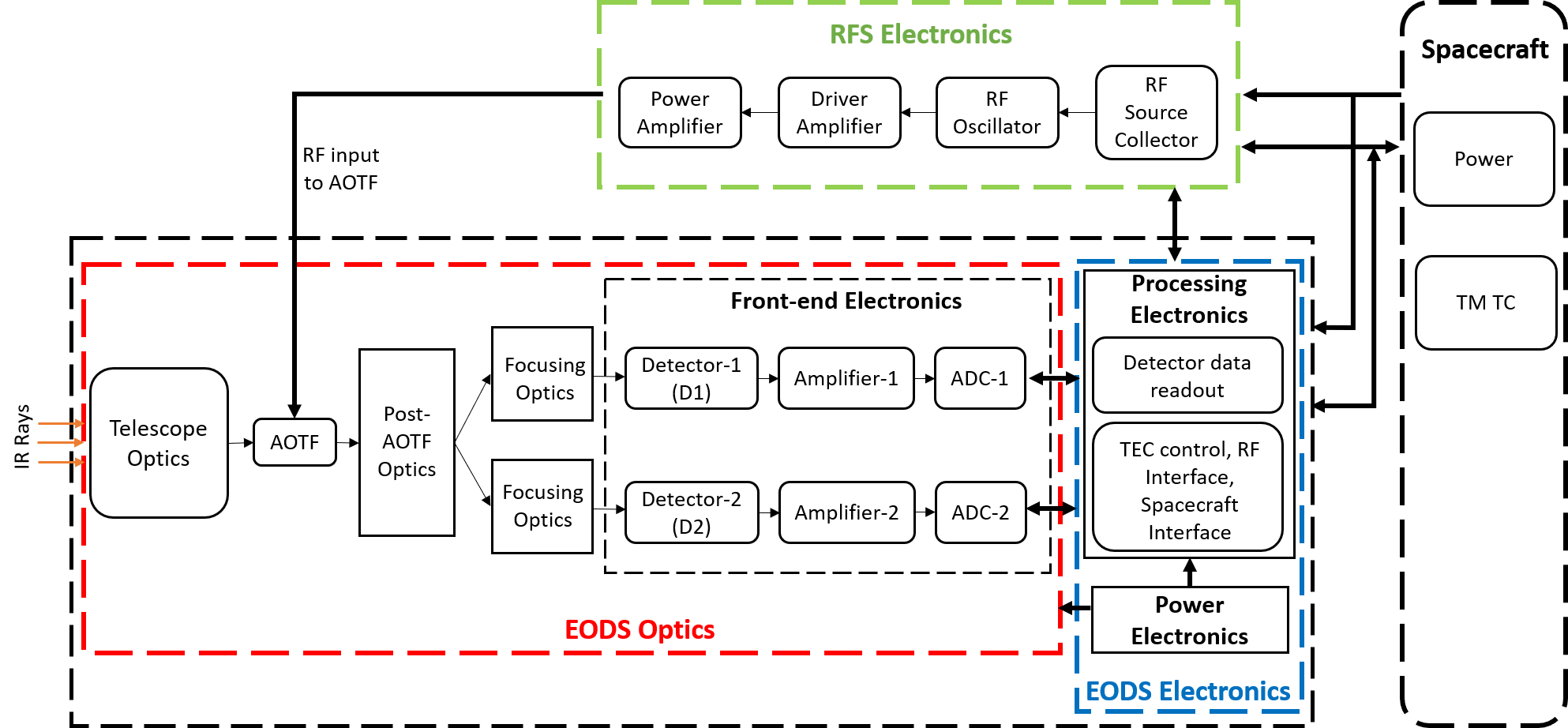}
\caption{The overall block diagram of the SHAPE instrument. All major components are shown. Refer to the text for details.}
\label{fig:blockdia}
\end{figure}

\begin{figure}[h!]
\centering
\includegraphics[scale=0.35]{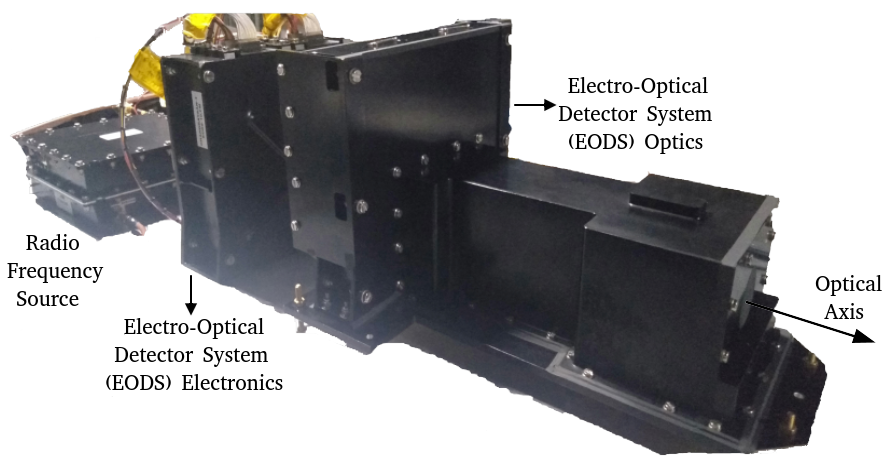}
\caption{Flight model of the SHAPE instrument. All three sub-systems are shown. The direction of the FOV/Optical axis is marked.}
\label{fig:shape_fm}
\end{figure}

To accomplish the above mentioned scientific objectives, the SHAPE instrument was designed using three packages: Electro-Optical Detector System (EODS)-Optics, EODS-Electronics and Radio Frequency Source (RFS). The overall block diagram of the instrument is shown in Figure \ref{fig:blockdia} and the flight model is shown in Figure \ref{fig:shape_fm}. The disc-integrated signal is collected, detected and digitized in the EODS-Optics package. The EODS-Electronics package houses the Processing Electronics (PE) and Power Distribution Electronics (PDE). To drive the AOTF, the RFS package generates RF power in the operating frequency range using an RF synthesizer, a driver amplifier and a power amplifier. Design and functionality of the packages are described in detail on the following sections and the instrument specifications are summarized in Table \ref{table:specs}.

\subsection{EODS-Optics}
EODS-Optics package houses the optics, opto-mechanical assembly, AOTF, detectors and the Front-end Electronics (FE) to process the detector signals (see Figure \ref{fig:blockdia}). The input signal is collected using the telescope optics and fed to the AOTF which generates two output beams. These beams are then focused onto two detectors. The signal from these detectors is amplified and pre-processed using the FE Electronics. Each of the major components is described in the subsequent sections.

\subsubsection{Filter and Detectors}
\begin{figure}
\centering
\includegraphics[scale=0.55]{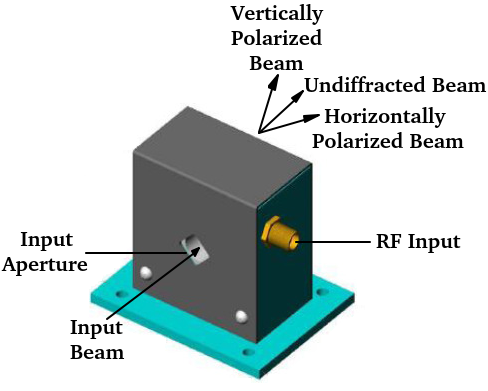}
\hspace{15mm}
\includegraphics[scale=0.5]{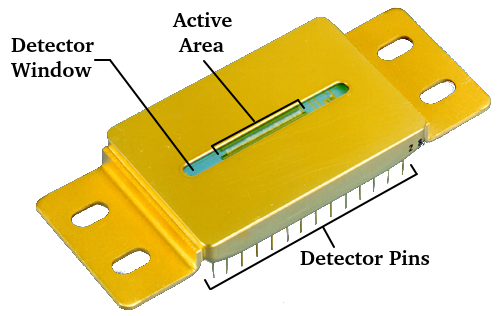}
\caption{\textit{Left: }AOTF manufactured by Brimrose Technology Corporation\protect\footnotemark[2]. The input aperture is shown along with the input beam and the three output beams. \textit{Right: }Linear array InGaAs detector developed by Sensors Unlimited\protect\footnotemark[3] with the detector pins, active area and window marked.}
\label{fig:aotf_det}
\end{figure}

In order to select the desired wavelength in the NIR band, an AOTF, manufactured by Brimrose Technology Corporation\footnotemark[2], is used. In the AOTF, a piezo-electric transducer is attached to a TeO$_2$ birefringent crystal to generate acoustic waves that form a spatial diffraction grating inside the crystal. Based on the principle of Bragg’s diffraction inside the crystal, the crystal acts as a tunable filter, tuned by an external RF source (\citealp{1969JOSA...59..744H, 1975ElL....11..617C}). The incident ray is split into two orthogonal linearly polarized beams known as ordinary (o) and extra-ordinary (e) beams having their wavelength set by the RF value \citep{1997OptLT..29..267G}. The AOTF along with the input and output beams is shown in Figure \ref{fig:aotf_det} (left). The spectro-polarimetric performance and characterization of an AOTF is described by \cite{2013ASInC...9..101N}, \cite{2015ExA....39..445A} and \cite{2022JATIS...8d4007J}. 
\footnotetext[2]{\url{https://www.brimrose.com/free-space-ao/acousto-optic-tunable-filters}}

The detectors chosen are InGaAs (Indium-Gallium-Arsenide) based pixelated linear arrays from Sensors Unlimited\footnotemark[3] shown in Figure \ref{fig:aotf_det} (right). The linear array consists of 256 pixels having a pixel size of $50\times500$ \textmu m$^2$. The photodiodes in each pixel are bonded with an associated Readout Integrated Circuit (ROIC), sensitive to the 0.9 to 1.7 \textmu m wavelength range and is cooled via a two-stage thermoelectric cooler (TEC). The ROIC is an active pixel device in which the photo-current generated by incident NIR radiation is buffered, amplified and stored, with serial readout. The array and ROIC are housed in a hermetically sealed package filled with dry Nitrogen. The NIR light focused onto the detectors is integrated on the detector and the time of integration is controlled by the Processing Electronics (PE).
\footnotetext[3]{\url{https://www.sensorsinc.com/products/focal-plane-arrays}}

\begin{figure}[h!]
\centering
\includegraphics[scale=0.55]{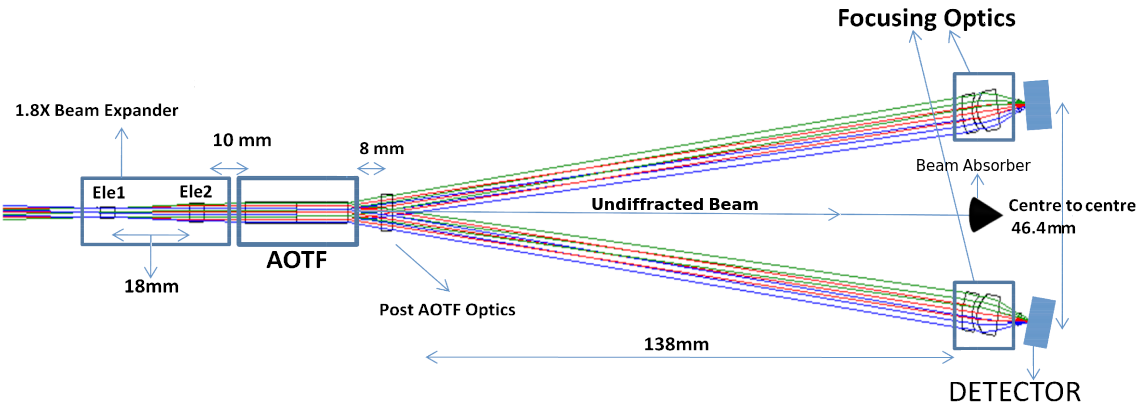}
\caption{The optics design of the instrument. All the major components in the EODS-Optics package along with the ray diagrams are shown.} 
\label{fig:optics_ray}
\end{figure}

\subsubsection{Optical Assembly}
In order to observe the full Earth disc from the LO, the instrument has a FOV of $\sim2.6^\mathrm{o}$, the input aperture at the front of the instrument is $\sim 2$ mm. The signal input on the instrument is collected and input to the AOTF, having an input aperture of $\sim 5$ mm. This is done using a beam expander with a power of $1.8\times$ to control the divergence and maintain the beam diameter at the AOTF input aperture. The output of the beam expander is the input of the AOTF. At the output, Post-AOTF Optics is used to maintain the divergence of each diffracted beam. This is a simple plano-convex lens whose diameter is higher than the AOTF output aperture in order to accommodate the diffracted beam for the whole bandwidth. After Post-AOTF optics, the beam is diffracted into three different beams: undiffracted beam and two diffracted beams. A pair of Focusing Optics is placed before the detector plane, whose orientation depends on the angle of diffraction and the position of the central wavelength at the detector plane. Each unit of Focusing Optics consists of two lenses: a concavo-convex focusing lens and an aspheric lens. The ray diagram along with all the optical components is shown in Figure \ref{fig:optics_ray}. The angle of diffraction between two beams is dependent on the wavelength and for a wavelength range of $1.0 - 1.7$ \textmu m, the angular separation varies from $8.24^\mathrm{o}-8.04^\mathrm{o}$. The design has been optimized for the central wavelength ($1.4$ \textmu m). The effective focal length of the instrument is $8.35 \pm 0.05$ mm. 

\subsubsection{Front-end Electronics}
\begin{figure}[h!]
\centering
\includegraphics[scale=0.5]{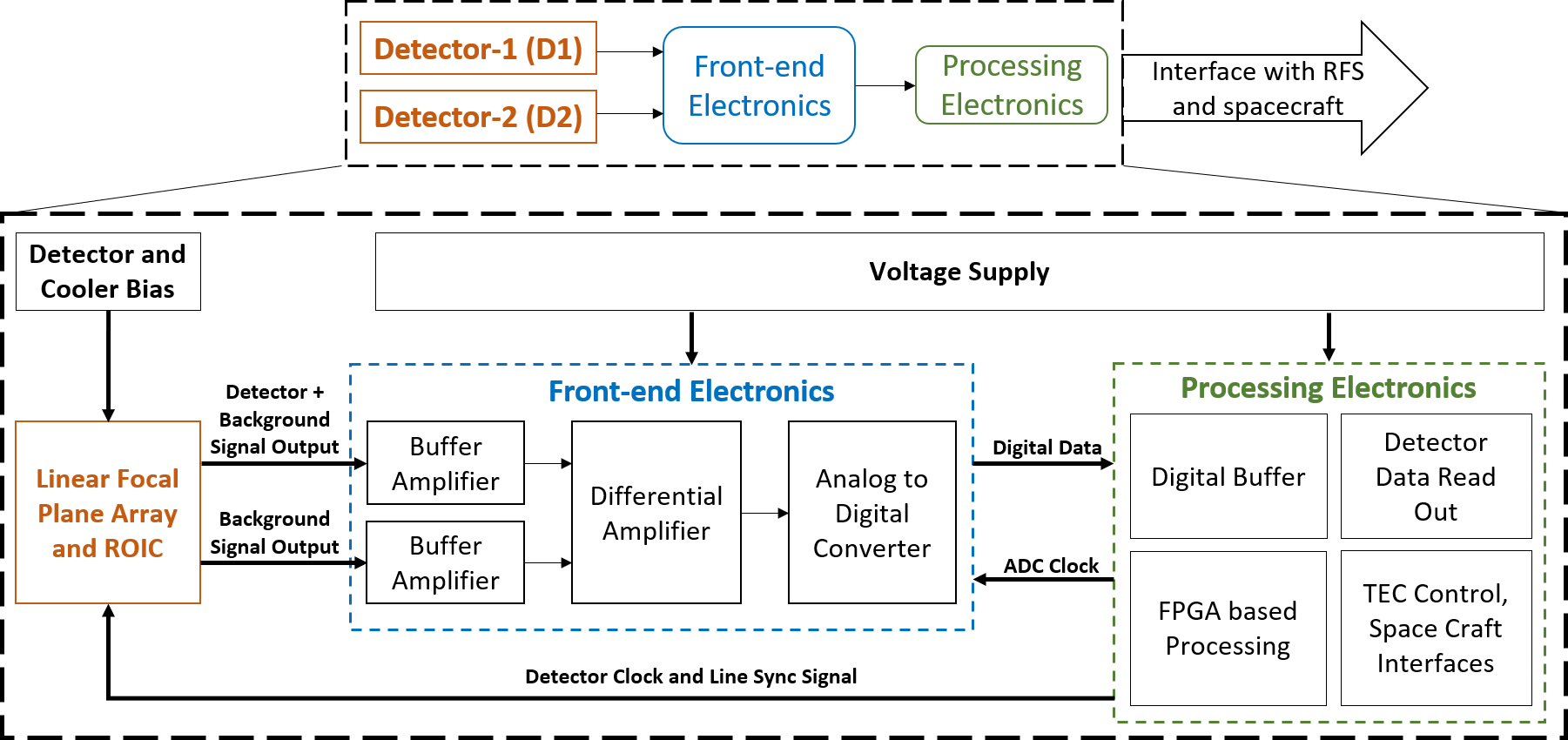}
\caption{Block diagram of Front-end Electronics. Data acquired by the detectors is sequentially processed by the Front-end and Processing Electronics. The components within each of these sub-systems are shown in the lower box and the flow is shown. All the electronics are powered by the Voltage Supply which is managed by the Power Distribution Electronics.}
\label{fig:eods_elec_block}
\end{figure}

FE includes the detectors and the electronics which amplify the signal from detectors, digitize it and send it to processing electronics for further processing. The schematic of the FE electronics is shown in Figure \ref{fig:eods_elec_block}. The NIR light is integrated on the detector, with integration time controlled by the PE. Each pixel provides two outputs via the in-built ROIC: (1) background signal and (2) detector signal plus background signal. Each detector has 256 pixels, and the two outputs are read serially using a clock. These outputs are individually buffered with low-noise, low-offset buffers and then differentially amplified to subtract the background using a low-noise differential operational amplifier. The subtraction is carried out via correlated double sampling which eliminates most of the integrator reset switching noise. The signal is then digitized using a 12-bit Analog to Digital Converter (ADC) which is sent to the processing electronics. The output is in ADC bins or Analog to Digital Units (ADU). 

\subsection{EODS-Electronics}
EODS-Electronics comprises of the PE and PDE modules as shown in Figure \ref{fig:blockdia}. The PE has been designed using an Actel RTAX2000S-1 Field Programmable Gate Array (FPGA). The PE module decodes all the commands and accordingly data is recorded, stored and played back. It also interfaces with the RFS package and sends it the start, stop and step frequency values along with the output RFS power which is to be fed to the AOTF. As there is no onboard recorder in the PM, a provision of 17 kByte memory in the FPGA to store the data packets is made. These data packets comprise of the data for the two detectors and housekeeping data, which includes the commanded frequency parameters, temperature of different components, etc. Along with the detector data, it stores the time, temperature of various components in the memory and makes 200-byte data packets which are sent via telemetry. SHAPE instrument has two modes of operation: Individual pixel and Sum modes, and the same can be selected by the PE. In the Individual pixel mode detector data is written individually for the specified number of pixels while in the Sum mode, the individual values for all the pixels are multiplied by the respective pixel gain values, summed and stored. The other module, PDE, houses two small Dual Output DC-DC Converters and associated electronics to cater to the power requirements of the EODS-Optics and EODS-Electronics packages.

\begin{table}[h!]
\begin{center}
\caption{Overall instrument specifications of the SHAPE Payload.}
\label{table:specs}
\begin{tabular}{|l|l|l|}
\hline
\textbf{Component} & \textbf{Parameter} & \textbf{Instrument Specification} \\
\hline
\multirow{4}{*}{Optics} 
    & Input Aperture Diameter      & 2 mm \\
    & FOV                         & $\sim2.6^\mathrm{o}$ \\
    & Effective Focal Length      & $8.35 \pm 0.05$ mm \\
\hline
\multirow{7}{*}{Filter}
    & AOTF Crystal                & TeO$_2$ \\
    & AOTF Aperture               & 5 mm $\times$ 5 mm \\
    & Spectral Range              & 1.0--1.7 \textmu m \\
    & Spectral Resolution         & 2--4 nm \\
    & RF Range                    & 80--135 MHz \\
    & RF Step Size                & Nominal: 200--2000 kHz (100 kHz steps) \\
    &                             & Calibration: 10 kHz and 20 kHz \\
    & RF Power                    & 0.5--2.0 W \\
    & Polarimetry                 & Linear polarisation \\
\hline
\multirow{5}{*}{Detector}
    & Detector Type               & InGaAs Linear Array \\
    & Total number of pixels      & 256 \\
    & Pixel size                  & $50 \times 500$ \textmu m$^2$ \\
    & Number of pixels recorded   & 10 (maximum), variable start pixel \\
\hline
\multirow{1}{*}{Detector Electronics}
    & Detector+Electronics Noise  & $<2.6$ mV or $<2$ ADC bins/ADU \\
    & Integration Time            & 10, 20, 50, 100, 200, 500, 1000 ms \\
    & Modes of operation          & Individual Pixel Mode: $1-10$ pixel data is independently recorded	\\
    &                             & Sum Mode: $2-10$ pixel data is summed and recorded	\\
\hline
\multirow{1}{*}{Data}
    & Data Volume                 & 17 kByte (allocated within the FPGA)\\
\hline
\multirow{4}{*}{Power}
    & EODS-Optics                 & 1.5 W \\
    & EODS-Electronics            & 9.8 W \\
    & RFS-Electronics             & 14.8 W \\
    & Total                       & 26.2 W \\
\hline
\multirow{3}{*}{Mechanical Dimensions}
    & EODS-Optics                 & $311.5 \times 215 \times 157$ mm$^3$ \\
    & EODS-Electronics            & $73 \times 183 \times 155$ mm$^3$ \\
    & RFS-Electronics             & $160 \times 146 \times 70$ mm$^3$ \\
\hline
\multirow{1}{*}{Instrument Mass}
    & Total                        & 4.8 kg (including thermal elements) \\
\hline
\end{tabular}
\end{center}
\end{table}

\subsection{RFS-Electronics}
SHAPE payload uses AOTF for filtering the NIR band ($1.0 - 1.7$ \textmu m) and the passband of the AOTF is determined by the frequency of the RF signal. For the given wavelength range, the RFS-Electronics package is designed to generate frequencies ranging from $80- 135$ MHz with power output in the range of $0.5 - 2.0$ W. This package consists of an RF Synthesizer Module and an Amplifier Module as shown in Figure \ref{fig:blockdia}. The RF Synthesizer generates the desired frequency sweep using a reference frequency signal derived from a crystal oscillator. The frequency of the RF synthesizer signal is set in real time based on the command data input received from EODS-Electronics. The RF synthesizer is controlled by an Actel RTAX-2000S FPGA with an embedded microprocessor. The FPGA-based controller is designed to receive the serial command from EODS-Electronics to extract the start frequency and step size, compute the corresponding register values for each sweep step, program the frequency synthesizer, and transmit housekeeping (HK) data back to EODS-Electronics through a serial interface. The output power of the RF Synthesizer is amplified to the desired value by using a driver amplifier and a power amplifier. The output of the RF amplifier is fed to the AOTF as shown in Figure \ref{fig:blockdia}. Along with the synthesizer and amplifier, the RFS package also consists of a DC-DC converter and associated relay circuitry for switching ON/OFF the converter. 

\section{Configuration of Mechanical and Thermal Design}
SHAPE is a low-weight instrument having a mass of less than 5 kgs, where in addition to the compact components such as the AOTF, the entire mechanical configuration is compact. The instrument also has various thermally critical components, especially in the EODS-Optics package where the temperatures have to be maintained within narrow ranges in the adverse lunar orbital conditions. 

\begin{figure}[h!]
\centering
\includegraphics[scale=0.4]{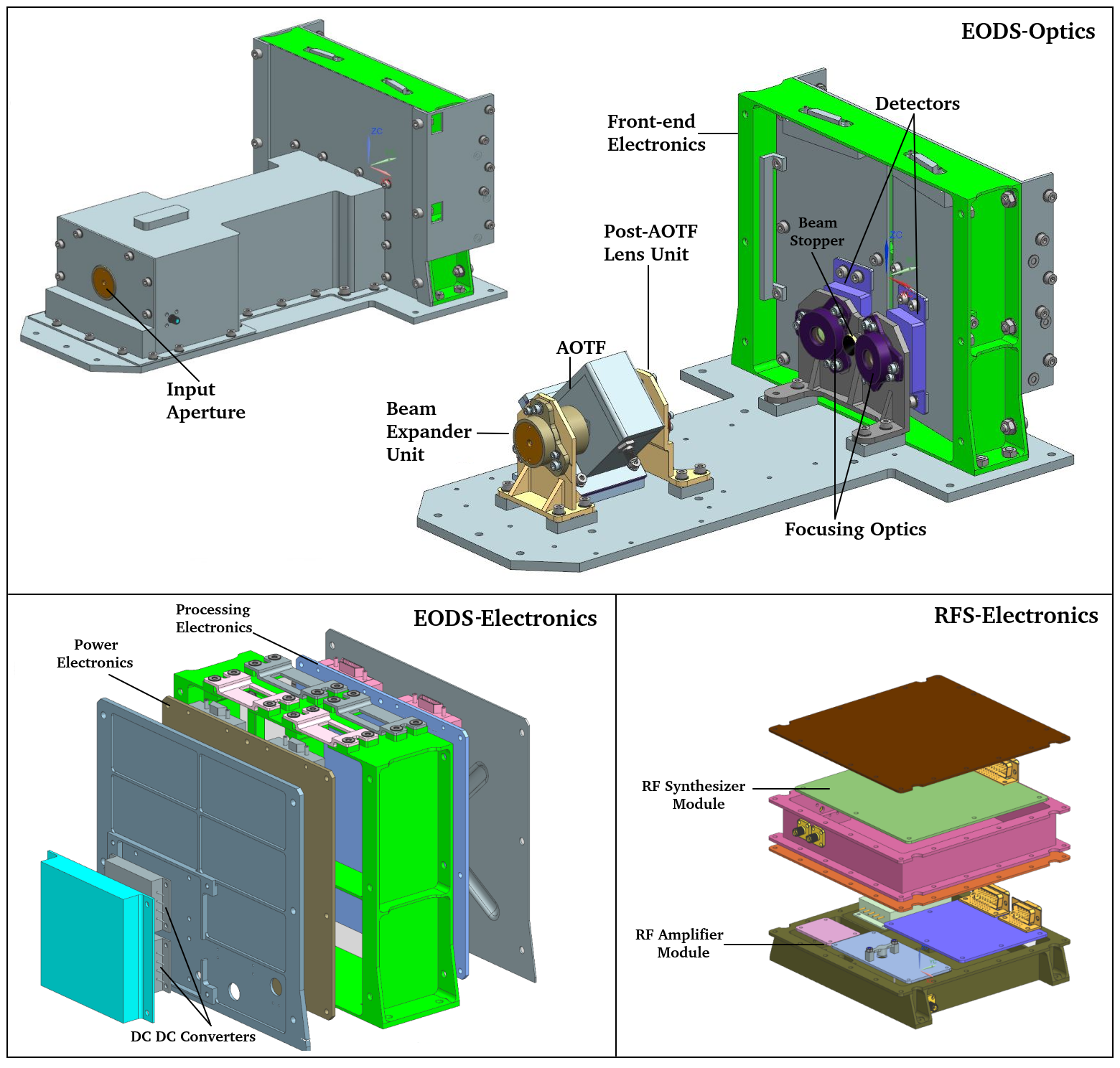}
\caption{Mechanical configuration of the EODS-Optics, EODS-Electronics and RFS-Electronics packages is shown with all major components.}
\label{fig:mech_cfg}
\end{figure}

\subsection{Mechanical Configuration}
Mechanically, SHAPE is configured into three units: EODS-Optics, EODS-Electronics and RFS-Electronics. The EODS-Optics unit has the optics, AOTF and opto-mechanical assemblies mounted on the optical bench. The lenses are provided with flanges that mount onto the cell stand and M2 tapped holes are provided near the mounting locations for alignment purposes. The detector units and corresponding Front-end Electronics are housed in a tray, which is vertically mounted on the optical bench as shown in the top panel of Figure \ref{fig:mech_cfg}. The detectors on the PCBs are mounted perpendicular to the incoming beam from the AOTF. The critical part of the alignment process is to focus the beam within the 500 \textmu m pixel (lateral direction). To facilitate alignment along the 0.5 mm axis of the detector, the detector mounting PCB has been designed with oval-shaped mounting holes, allowing controlled lateral movement. Additionally, to enable independent adjustment of the detector PCBs, the mechanical housing of the EODS package incorporates a provision for fine movement using adjustable screws. The EODS-Electronics unit has the PE and PDE PCBs stacked and mounted vertically (shown in the bottom-left of Figure \ref{fig:mech_cfg}). The RFS unit has two trays stacked together to house the synthesizer, power amplifier, driver amplifier and interface card including one DC-DC converter (shown in the bottom-right of Figure \ref{fig:mech_cfg}). The optical bench, electronics trays and the covers are all made of Al-6061. The overall three-dimensional configuration of the three packages is shown in Figure \ref{fig:mech_cfg}. To ensure the structural integrity of the mechanical design, normal mode and quasi-static stress analysis were also performed.

\begin{figure}[h!]
\centering
\includegraphics[scale=0.54]{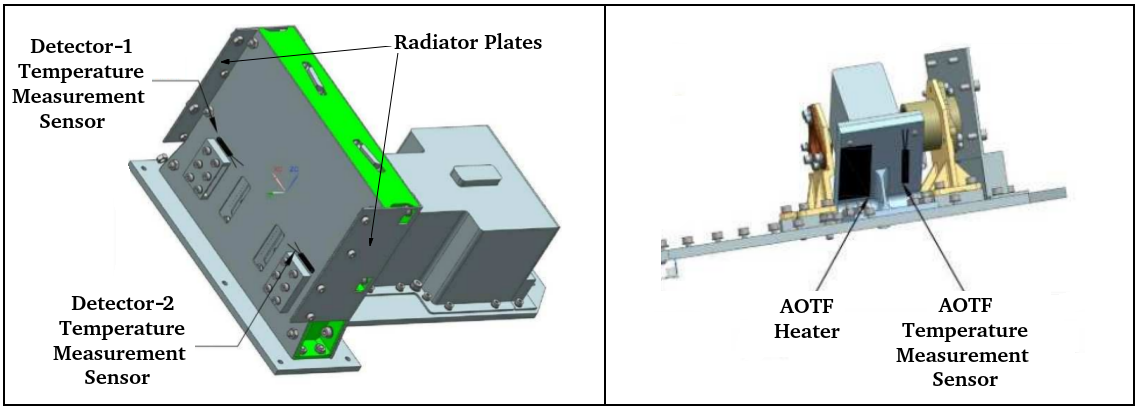}
\caption{Thermal design configuration in EODS-Optics package showing the placement of the thermal elements: heater and radiators. The figure also shows the locations where temperature measurements are made for the detectors and the AOTF.}
\label{fig:ther_cfg}
\end{figure}

\subsection{Thermal Design}
SHAPE payload is designed with various specialized sub-units: InGaAs detectors, AOTF and optics. It is necessary to maintain the temperatures of the sub-units as well as the EODS-Electronics and RFS packages at all times when the spacecraft is in orbit for the optimal performance of the instrument. Along with the external heat loads from Sun, Earth and Moon, there is an internal heat dissipation of $\sim 15$ W from the three packages. The various levels of temperatures, as listed in Table \ref{table:etls}, have to be maintained during SHAPE operations in both LO and HEO for the optimum performance of the instrument. The ‘Min’ and ‘Max’ temperatures corresponding to ‘Turn On’ and ‘Thermal Design’ give the temperature range within which all the sub-units need to be maintained throughout the mission life for switching on and operating the instrument respectively. An efficient thermal control system was designed by using electrical heaters, thermal radiators, PGS flexible thermal straps, Multi Layered Insulation (MLI) blankets, thermal interface materials and thermistors to maintain the temperatures of all the sub-units within the specified range during cold and hot environments in the orbit. To achieve the specified temperatures, the EODS-Optics package is conductively and radiatively isolated from the spacecraft deck using 7 mm spacers. To maintain the temperatures of the detectors, thermal interface material was used below the detectors. Two sets of PGS flexible thermal straps were employed to transfer the internally generated heat from detectors to thermal radiators. This maintained the detector housing temperature between 17-23$^\mathrm{\circ}$C and the detector was further cooled via the in-built thermo-electric cooler (TEC) to provide a $\triangle\mathrm{T}\sim 20^\mathrm{\circ}$C $\pm 1^\mathrm{\circ}$C. AOTF was mounted with compressible PGS thermal interface material and augmented with heaters. The EODS-Electronics and RFS-Electronics packages were flush mounted on the spacecraft panel using thermal interface material. All three packages were completely covered with MLI insulation blankets except for their mounting interfaces, Optics opening and thermal radiators. There were a set of thermistors which have been bonded at various locations in the instrument to continuously monitor the temperatures of all the critical components and provide feedback for thermal management. The thermal design configuration for the EODS-Optics package is shown in Figure \ref{fig:ther_cfg}. Based on the thermal analysis for the lunar orbit, the payload is operable in the orbit segments beyond Noon-Midnight $\pm 60$ days (i.e., during the regions of the lunar orbit situated outside of the $\pm 60$ day window relative to Noon-Midnight). Within the Noon-Midnight $\pm 60$ day region, the payload is limited to operation only on the dark side of the Moon during each orbit. The thermal management system was also able to successfully maintain the temperatures of the components in the HEO around Earth.

\begin{table}
\begin{center}
\caption{Temperature limits for various packages in the SHAPE payload.}
\label{table:etls}
\begin{tabular}{|c|c|c|cc|cc|}
\hline
Sl. No. & Package       & \begin{tabular}[c]{@{}c@{}}Component\end{tabular} & \multicolumn{2}{c|}{\begin{tabular}[c]{@{}c@{}}Turn On ($^\mathrm{o}$C)\end{tabular}} & \multicolumn{2}{c|}{\begin{tabular}[c]{@{}c@{}}Thermal Design ($^\mathrm{o}$C)\end{tabular}} \\ \hline
     &								&                                               & \multicolumn{1}{c|}{Min}                                & Max                               & \multicolumn{1}{c|}{Min}                                     & Max                                   \\ \hline
1    & EODS-Optics & -                                                & \multicolumn{1}{c|}{-}                                & -                               & \multicolumn{1}{c|}{0}                                     & 40                                   \\ \hline
2    & EODS-Electronics   & -                                                & \multicolumn{1}{c|}{-25}                              & 55                              & \multicolumn{1}{c|}{-10}                                   & 40                                   \\ \hline
3    & RFS-Electronics         & -                                                 & \multicolumn{1}{c|}{-15}                              & 55                              & \multicolumn{1}{c|}{-10}                                   & 40                                   \\ \hline
4    & \multirow{3}{*}{\begin{tabular}[c]{@{}c@{}}EODS-Optics\\ Components\end{tabular}} & AOTF                                                & \multicolumn{1}{c|}{22}                               & 28                              & \multicolumn{1}{c|}{22}                                    & 28                                   \\ \cline{1-1} \cline{3-7} 
5    &                                                                            & Detector                                                & \multicolumn{1}{c|}{17}                               & 23                              & \multicolumn{1}{c|}{17}                                    & 23                                   \\ \cline{1-1} \cline{3-7} 
6    &                                                                            & Optics                                                 & \multicolumn{1}{c|}{17}                               & 27                              & \multicolumn{1}{c|}{17}                                    & 27                                   \\ \hline
\end{tabular}
\end{center}
\end{table}

\section{Ground Test and Performance}
During the development and realization of the payload various tests were carried out at different stages to ensure that the payload performance meets the specifications required to achieve the science goals. These tests included the optical alignment, detector characterization and the EODS-RFS Interface tests. After the functional verification of the full system, various environmental tests were conducted simulating the launch and space environments. The full instrument was then calibrated for accurate onboard measurements.

\subsection{Developmental Tests}
The SHAPE instrument was developed in a time span of 3 years. During the development of the payload, each subsystem was individually tested and later all the sub-systems were interfaced together, post which end-to-end testing was carried out. In the instrument, all the parameters such as the integration time, spectrum start and stop frequency, spectrum step frequency, etc. are variable and can be set via telecommand as per the requirement of observation. Each of these telecommands for both modes of operation (Individual Pixel Mode and Sum Mode) were also tested at every step of the realization of the payload. 

\subsubsection{Detector Characterization}
\begin{figure}
\centering
\includegraphics[scale=0.43]{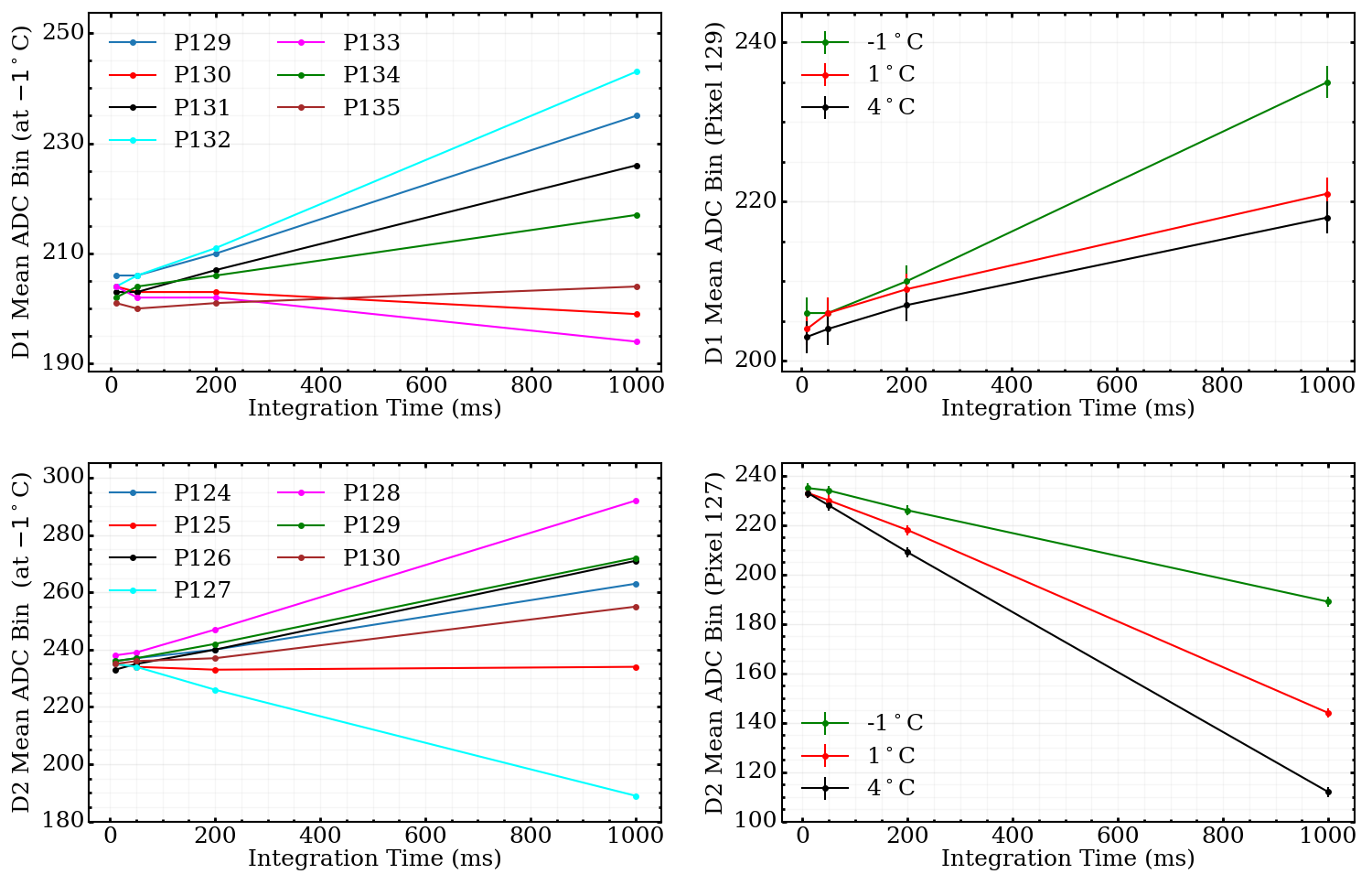}
\caption{(left) Detector dark signal variation with integration time for different pixels of Detector-1 (top) and Detector-2 (bottom). (right) Variation of detector dark signal with integration time at different detector temperatures for Pixel 129 of Detector-1 (top) and  Pixel 127 of Detector-2.}
\label{fig:det_char}
\end{figure}

After the optical alignment, the central pixels of the detector were characterized for their dark current with integration time and temperature. Figure \ref{fig:det_char} shows the variation of dark signal (in terms of ADC bins) for both detectors. The photodiodes in the detector are nominally held at zero bias to minimize dark current. But due to non-uniformities in the readout, there can be an offset of $\pm3$ mV. As the input offset voltage may be positive or negative, and the dark current is equal to the input offset voltage of that pixel divided by the shunt resistance, the dark current may also have either sign. Hence in Figure \ref{fig:det_char}, it is seen that with increasing integration time, the Mean ADC bin value either increases or decreases for different pixels. The dark signal magnitude also increases with an increase in the detector temperature. In each on-board observation, the background measurement would consist of the dark signal, which will be subtracted from the science observations.

\begin{figure}[h!]
\centering
\includegraphics[scale=0.45]{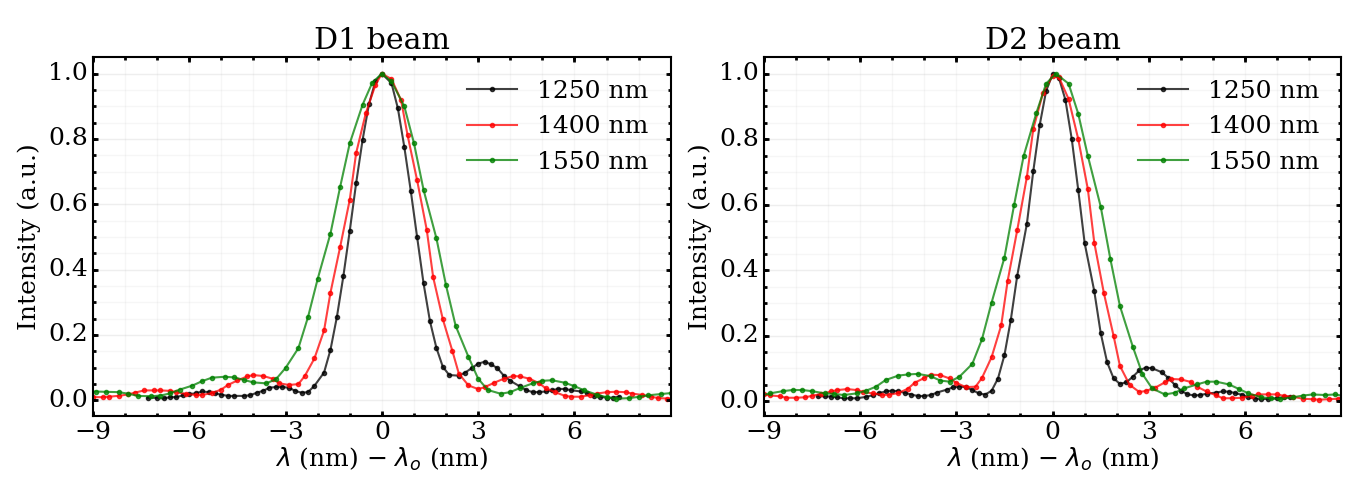}
\caption{AOTF transfer functions for both output beams at three wavelengths: 1250 nm, 1400 nm and 1550 nm. The wavelength ($\lambda$) for each transfer function is subtracted by the respective central wavelength ($\lambda_o$).}
\label{fig:aotf_tf}
\end{figure}

\subsubsection{AOTF Characterization}
The characterization of the AOTF was performed to determine its spectral transfer function over the spectral range of 1.0 to 1.7 \textmu m, covering the operational bandwidth of the instrument. The AOTF transfer function typically follows a $sinc^2$ profile, which describes the intensity distribution as a function of wavelength around the central tuning frequency. A comprehensive calibration methodology and discussion of the transfer function profile can be found in \cite{2015ExA....39..445A} and \cite{2022JATIS...8d4007J}. Measurements were conducted for both output beams produced by the AOTF: the ordinary and extraordinary beams. The transfer functions obtained for these two beams exhibit distinct characteristics due to the birefringent nature of the AOTF crystal. The AOTF transfer functions at three wavelengths are shown in Figure \ref{fig:aotf_tf}, which also indicates the variations in resolution with wavelength.

\subsubsection{Temperature Dependence Test of RFS}
\begin{figure}[h!]
\centering
\includegraphics[scale=0.44]{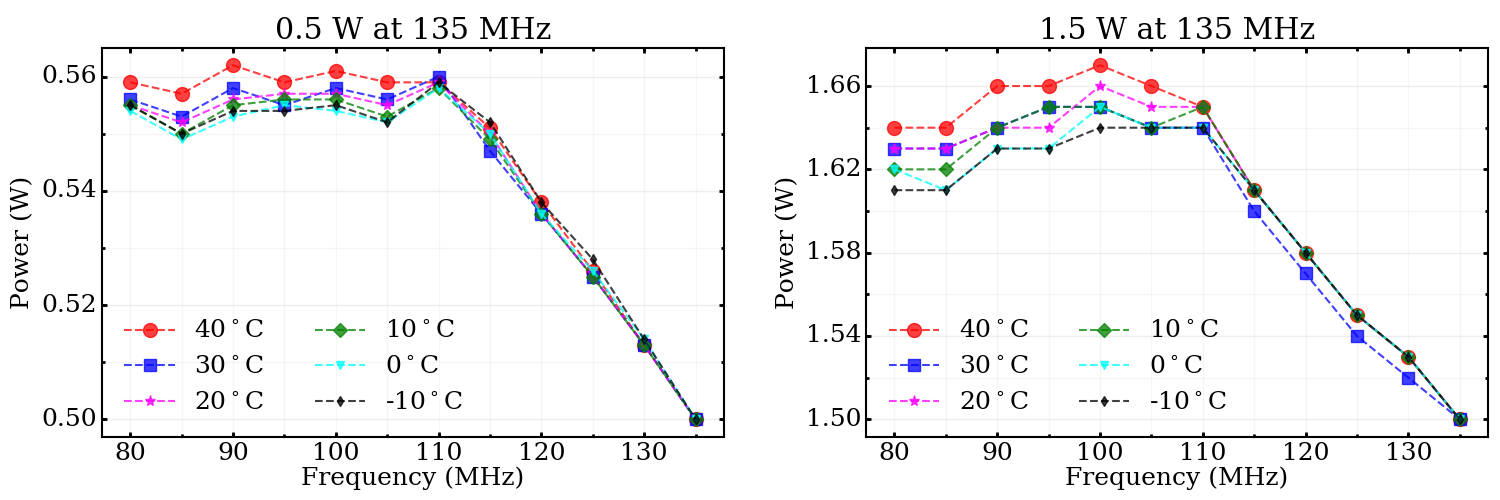}
\caption{Results from the RF temperature test for different power settings. The input power was set as 0.5 and 1.5 W at 135 MHz and the temperature of the RF system was varied from $-10^\mathrm{o}$C to $40^\mathrm{o}$C.}
\label{fig:rfs_temp}
\end{figure}

The RFS output power varies with frequency and the temperature of the power amplifier. To characterize the variation of output power at various temperatures, the RFS-Electronics package was placed in a temperature-controlled chamber and the temperature was varied from $-15^\mathrm{o}$C to $55^\mathrm{o}$C. The output power was measured using a Power Meter in the frequency range of 80 to 135 MHz. The results for 0.5 W and 1.5 W are shown in Figure \ref{fig:rfs_temp}. The variation in power at a frequency due to temperature is $<0.04$ Watts for the entire operating RF range. There is also a consistent trend in the variation of power with frequency. This is incorporated in the system response, which will be deconvolved to obtain the spectra. There is a continuous on-board measurement of the power amplifier temperature, which will enable the estimation of the RF power sent to the AOTF.

\subsubsection{EODS-RFS Interface Test}
The RFS-Electronics generates the RF signal at a given frequency and power when a command is sent from the PE card in EODS-Electronics. This RF signal is then fed to the AOTF. After all three packages were individually tested, EODS packages were interfaced with the RFS package and tested. During the test, all the voltage signals sent between the PE and RFS were verified. The power of the output RF signal is controlled by changing the voltage for a digital-to-analog converter (DAC). The power for different DAC voltages was measured to obtain the DAC voltage and output power look-up table in the operating range of $0.5-2.0$ W. Using a Krypton lamp as the input source, the spectrum was recorded at different powers for frequency step sizes of 20 kHz and 200 kHz. 

\subsection{Optical Alignment Test}
\begin{figure}[h!]
\centering
\includegraphics[scale=0.35]{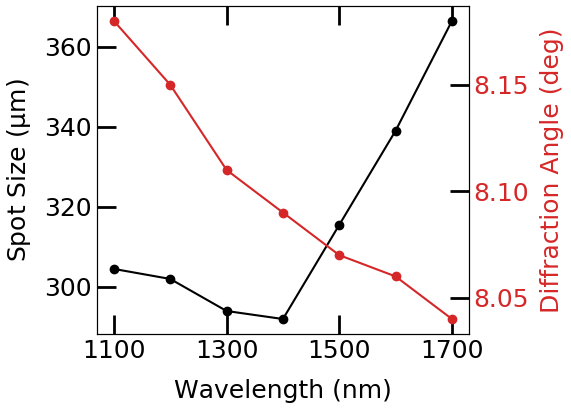}
\caption{Wavelength dependence of the spot size as per the optical design of the payload and the diffraction angle for the AOTF.}
\label{fig:blurdia}
\end{figure}

\begin{figure}[h!]
\centering
\includegraphics[scale=0.5]{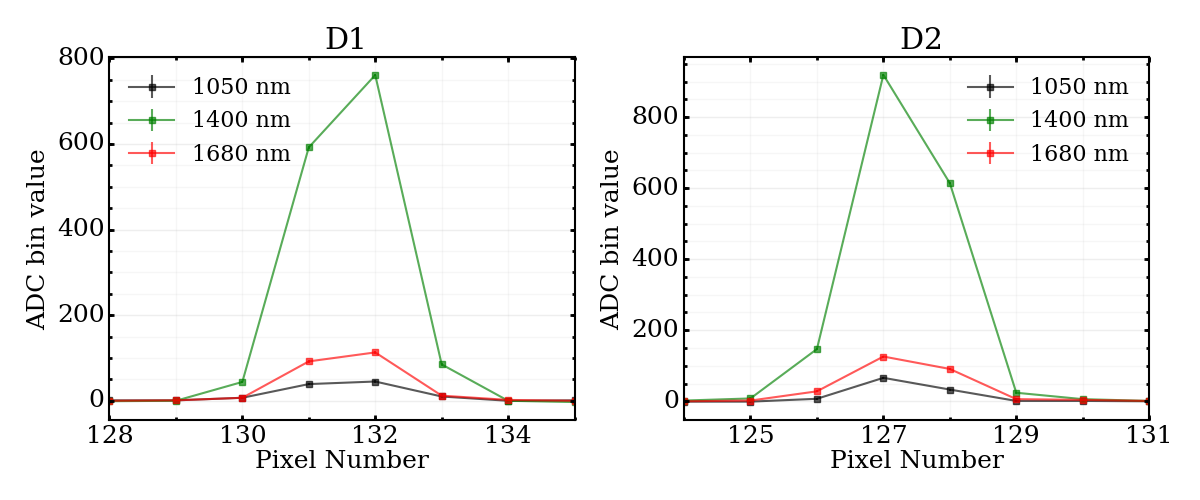}
\caption{Pixel-wise data recorded (on-axis) after the optical alignment of the instrument in both detectors, filtering at three wavelengths.}
\label{fig:align}
\end{figure}

The InGaAs detectors used in the instrument have 256 pixels which have a width of $500$ \textmu m in the horizontal direction and a length of $50$ \textmu m in the vertical direction. The optics is designed to ensure that the spot size (blur diameter) is $< 500$ \textmu m for the entire wavelength range and entire FOV. The blur size at each of the wavelengths depends upon the diffraction angle as well. Figure \ref{fig:blurdia} shows the variation of the spot size with wavelength as per the optical design. Hence, alignment of the instrument is crucial in order to avoid the loss of signal in the horizontal direction at all wavelengths throughout the operating range. On the vertical end, there is an array of 256 pixels, so there is no problem. To carry out the alignment, an un-polarized broadband source was used. The output wavelength used for alignment was set using the AOTF. The three lens assemblies were mounted in three steps and after verifying the output at each stage, shims were added to the sub-assembly wherever needed to correct for the tip-tilt errors. After completion, the final alignment of the system was verified at three different wavelengths as shown in Figure \ref{fig:align}. The central pixel numbers after alignment for Detector-1 and Detector-2 are 132 and 127, respectively. The achieved spot size for the operating range is $\leq 400$ \textmu m for on- and off-axis locations within the FOV, which is less than the pixel size in the horizontal direction. For off-axis object locations, the measured maximum vertical shift of the spot centroid is $\leq2$ pixels, resulting in a total spot spread of 7 pixels. Since the spot size remains consistently within 3 to 4 pixels, the horizontal pixel is sufficient to capture the signal for objects located anywhere within the FOV. The spot size is consistent during the in-orbit observations \citep{2024arXiv241207416N}.

\subsection{Environmental Tests}
It is standard practice in space instrumentation to subject the payload to a series of qualification and environmental tests that simulate the conditions expected during launch and in-orbit operation (LO and HEO). These tests are essential to ensure the structural integrity, functional reliability, and overall performance of the instrument under harsh space-like environments. SHAPE instrument was subjected to a comprehensive suite of environmental tests. These include thermo-vacuum, vibration and electromagnetic compatibility. Throughout each of these test phases, the instrument’s functionality and performance were verified to ensure that all subsystems operated within the expected parameters. The successful completion of these tests provides confidence that the SHAPE instrument is fully qualified for deployment in its intended space environment.

\subsubsection{Vibration Test}
\begin{figure}[h!]
\centering
\includegraphics[scale=0.378]{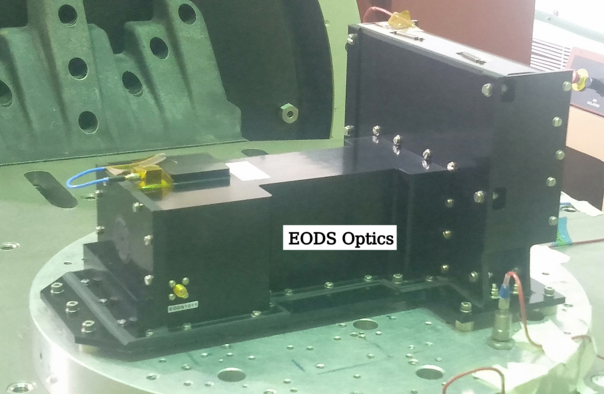}
\caption{The EODS-Optics package positioned on the vibration stage as part of the environmental testing. The subsystem is firmly clamped to the stage, and sensors are attached to monitor its dynamic response as the stage subjects it to the launch-level vibration environment.}
\label{fig:vib_emiemc}
\end{figure}

For every space-based instrument, vibration tests are carried out to demonstrate its ability to sustain the dynamic loads during launch and thereafter, without compromising on the structural integrity and the functionality of the instrument. Each package of the SHAPE instrument was subject to random and sine vibration normal and parallel to the mounting plane \citep{N2021CDR}. For the random vibration, the overall g$_\mathrm{RMS}$ is 9.9g and 7.9g for normal and parallel to the mounting plane cases. The vibration levels for sine vibration vary with frequency and range from 13.3g to 5.3g for frequencies ranging between 20 Hz and 100 Hz. Figure \ref{fig:vib_emiemc} shows the EODS-Optics package mounted on the vibration table. The instrument's performance was monitored pre- and post-vibration and the performance integrity was maintained. 

\subsubsection{Thermo-vacuum Test}
\begin{figure}[h!]
\centering
\includegraphics[scale=0.35]{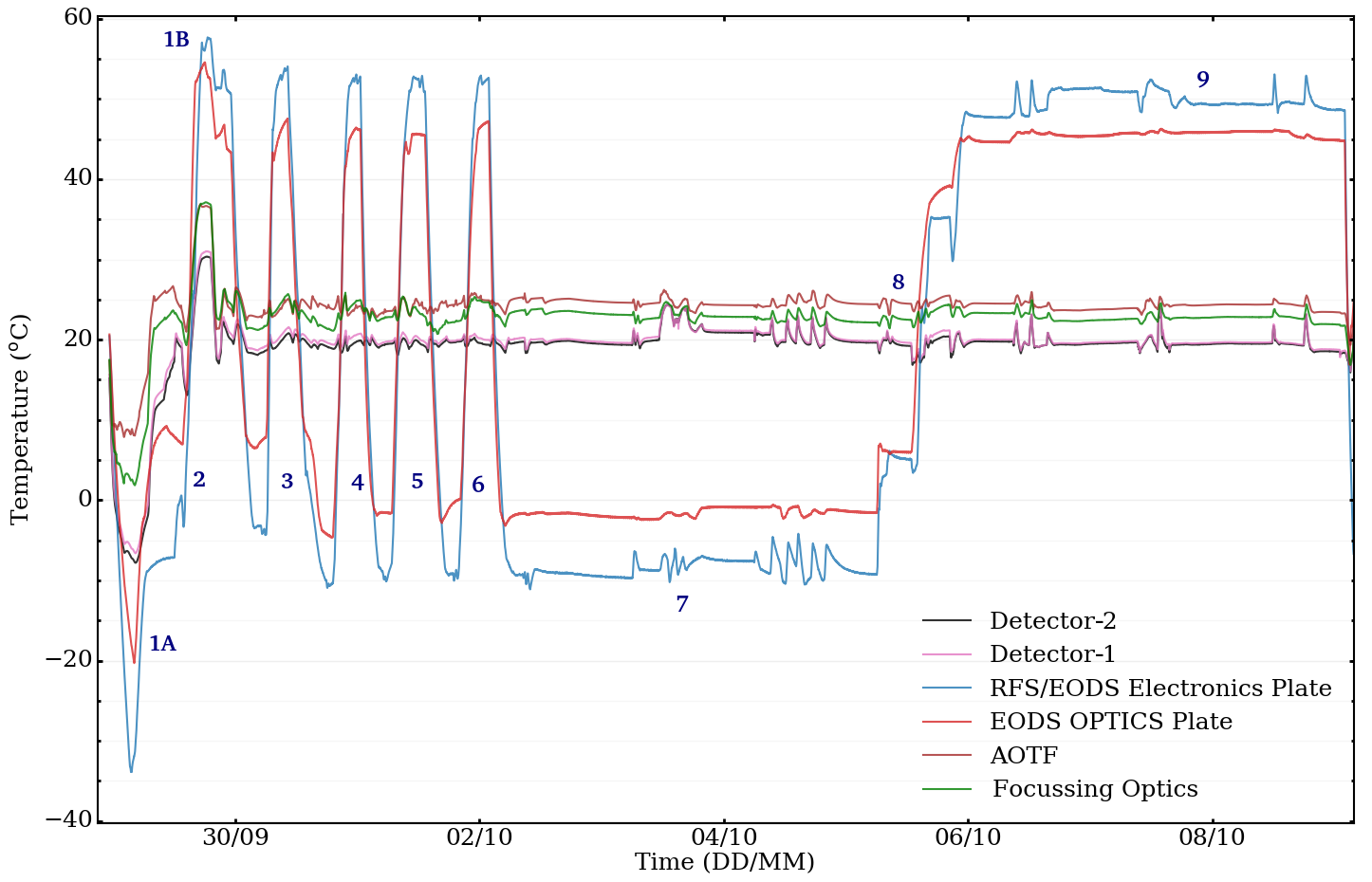}
\caption{SHAPE TVac Test profile. In stage 1, the instrument electronics and component temperatures were set to the minimum (1A) and maximum (1B) non-operating temperatures for two hours. Stages 2, 3, 4, 5 and 6 include five short cycles of two hrs duration each, where the temperatures were maintained within the qualification limits. Cold and Hot soak tests were carried out in stages 7 and 9 respectively, where the instrument components' temperatures were maintained within the Qualification limits. The total soak duration was 72 hours, out of which the instrument was switched on for 24 hours. The thermal balance test is conducted during stage 8 to validate the thermal design implementation.}
\label{fig:tvac_profile}
\end{figure}

\begin{figure}[h!]
\centering
\includegraphics[scale=0.45]{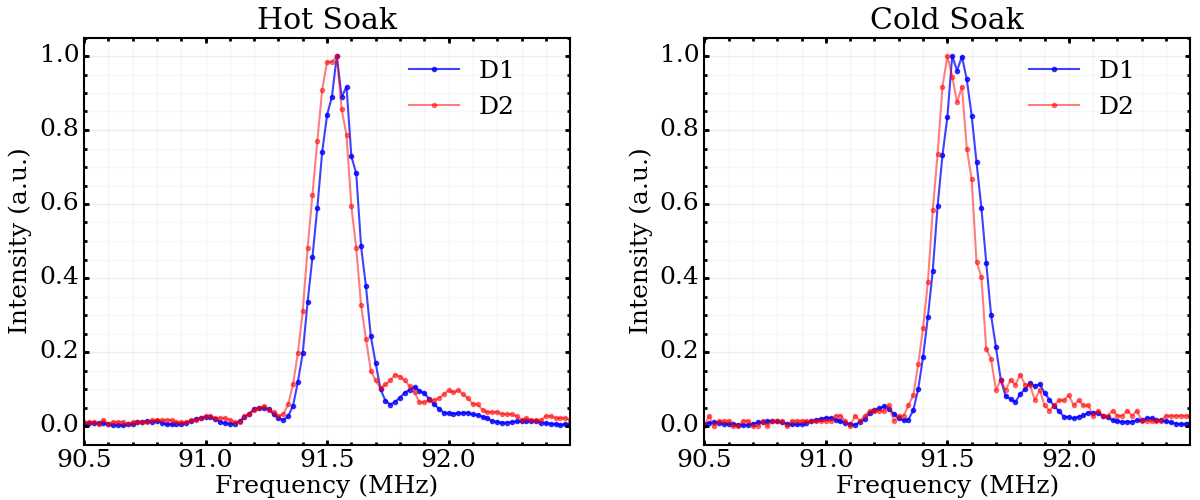}
\caption{SHAPE TVac Test results for hot and cold soaks. A LASER source at 1515 nm was used as the input to the instrument, the RF power input to the AOTF was 0.5 W and data was acquired for 10 ms integration time.}
\label{fig:tvac_sinc}
\end{figure}

To check the performance of the instrument in the vacuum of space across the entire range of temperatures (see Table \ref{table:etls}) in space, Thermo-vacuum (TVac) tests were carried out with the FM instrument along with the ground console system for commanding. The instrument was tested with a LASER (1515 nm) as the source and dark frames were also acquired to have an estimate of the background. Measurements were carried out in both modes: Individual Pixel and Sum. The entire functionality of the instrument was tested: variation of integration time and RF input power, and variation of the start, stop and step frequencies. The test results confirmed that the thermal design accurately reflects the intended specifications. The TVac test was carried out in various stages as shown in Figure \ref{fig:tvac_profile} \citep{N2021CDR}. The temperatures of the individual components like the AOTF and detectors were maintained manually using heaters during the test and this results in the minor variations seen in the measured temperatures. The AOTF temperature during the hot and cold soaks is maintained within the operational limits. Figure \ref{fig:tvac_sinc} shows the response function obtained at 1515 nm during the hot and cold soaks of TVac test. During the period in which the temperature is maintained, a significant change in spectral resolution was not observed, with variations remaining within $\pm0.2$ nm.

\subsubsection{Electromagnetic Induction (EMI) and Electromagnetic Conduction (EMC) Tests}
Electromagnetic interference and compatibility tests were carried out to assess the impact of electromagnetic interference phenomena between the various satellite subsystems. EMI/EMC tests include Conductive Emission (CE), Radiative Emission (RE), Radiation Susceptibility (RS) and Conductive Susceptibility (CS) tests  \citep{N2021CDR}. CE tests were carried out to measure the conducted emissions effect of the payload sub-systems switch-on and switch-off on the power line from the spacecraft. The measured values were below the specified limits, confirming that SHAPE operations do not cause any adverse impact on other spacecraft elements. For RE, the test was carried out for a large frequency range of $0.01-40000$ MHz, with the electric field limits varied based on the requirements and sensitivities of the other systems on the spacecraft. The measured instrument radiation was within the set limits. During the RS tests, SHAPE was tested for a field of 5V/m at varied frequencies corresponding to the other components spacecraft. During these tests, it was seen that the dark signals were varying on application of an external field. The other three tests had no effect on the system. To mitigate the effect of external field, all the harnesses, connectors and edges were covered with Copper tape and the RFS-Electronics package was enclosed in a Copper mesh. For CS tests, a noise of 1 V$_\mathrm{RMS}$ was injected in the power line from the spacecraft to verify its effect on the instrument. The instrument performance was nominal during these tests.

\subsection{Ground Calibration}
Ground calibration of the instrument includes radiometric, spectroscopic, polarimetric and field calibrations. The data obtained from each of these calibrations will be combined to form a Calibration Database, which will be used for deconvolving the instrument response to obtain the input signals. Each of these calibrations is described in the following subsections.

\subsubsection{Radiometric Calibration}
The output intensity from the instrument is in units of ADC bin values. For radiometric calibration of the detectors, the Krypton Lamp was placed at the same distance from a NIST-calibrated InGaAs detector\footnotemark[4] and the two InGaAs detectors of the SHAPE instrument. The power falling on the NIST calibrated detector \citep{Podobedov2016} is measured and then used to estimate the power falling on each pixel of the SHAPE detectors. Following this, the scaling factors were obtained to convert the ADC bin value to intensity. As the gain for each pixel is different, the scaling factors for all pixels are different. These scaling factors for Detector-1 and Detector-2 range between $(1.08-1.25)\times 10^{-15}$ W/s/ADC-channel and $(1.08-1.13)\times 10^{-15}$ W/s/ADC-channel, respectively. To account for the optics and AOTF ahead of the detectors, their respective transmission efficiencies are considered to obtain the final conversion factors.
\footnotetext[4]{\url{https://www.newport.com/c/standard-photodiode-sensors}}

\begin{figure}[h!]
\centering
\includegraphics[scale=0.43]{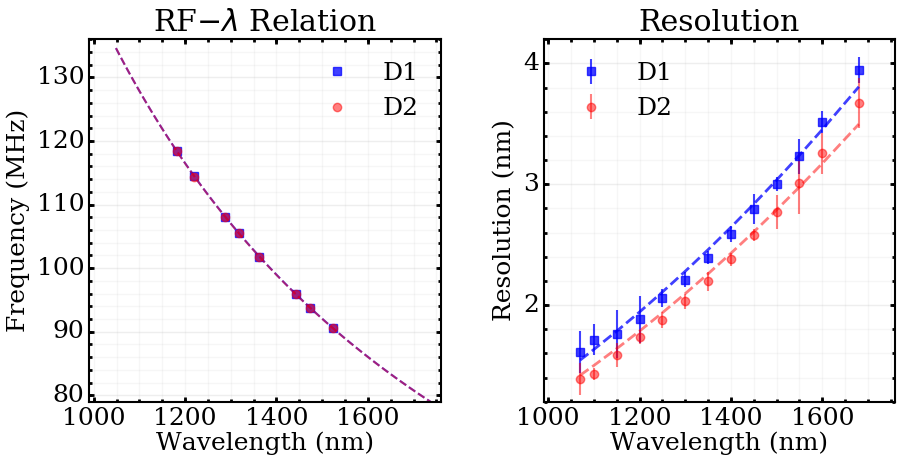}
\caption{Spectroscopic calibration results for SHAPE. (left) Frequency-wavelength relation for both detectors. (right) Variation of the spectral resolution of the SHAPE instrument with wavelength.}
\label{fig:spec_cal}
\end{figure}

\subsubsection{Spectroscopic Calibration}
Spectroscopic Calibration of the instrument includes estimation of the spectral response, spectral resolution and the frequency-wavelength tuning relation (see \citealp{2015ExA....39..445A,2022JATIS...8d4007J} for details). To obtain the frequency-wavelength or RF$-\lambda$ relation, a Krypton lamp having standard lines at fixed wavelengths was used. The wavelengths for these standard lines were compared with the frequency at which the line is seen and the RF$-\lambda$ relation was derived. The spectral response for the AOTF is a $sinc^2$. To obtain the spectral response, a broadband source along with a monochromator was used to provide a monochromatic input to the instrument and the response was recorded. This measurement was carried out in intervals of 50 nm on the operating wavelength range. The FWHM of the $sinc^2$ function is used to obtain the spectral resolution. The spectral resolution varies with wavelength as given by \cite{2015ExA....39..445A}. The results from the spectroscopic calibration are shown in Figure \ref{fig:spec_cal}. A significant difference in the spectral resolution is seen for the two output beams. The major reason for this difference is due to different interaction lengths between acoustic wave and light wave inside the crystal for the two beams \citep{2015ExA....39..445A}.

\subsubsection{Polarimetric Calibration}
\begin{figure}[h!]
\centering
\includegraphics[scale=0.39]{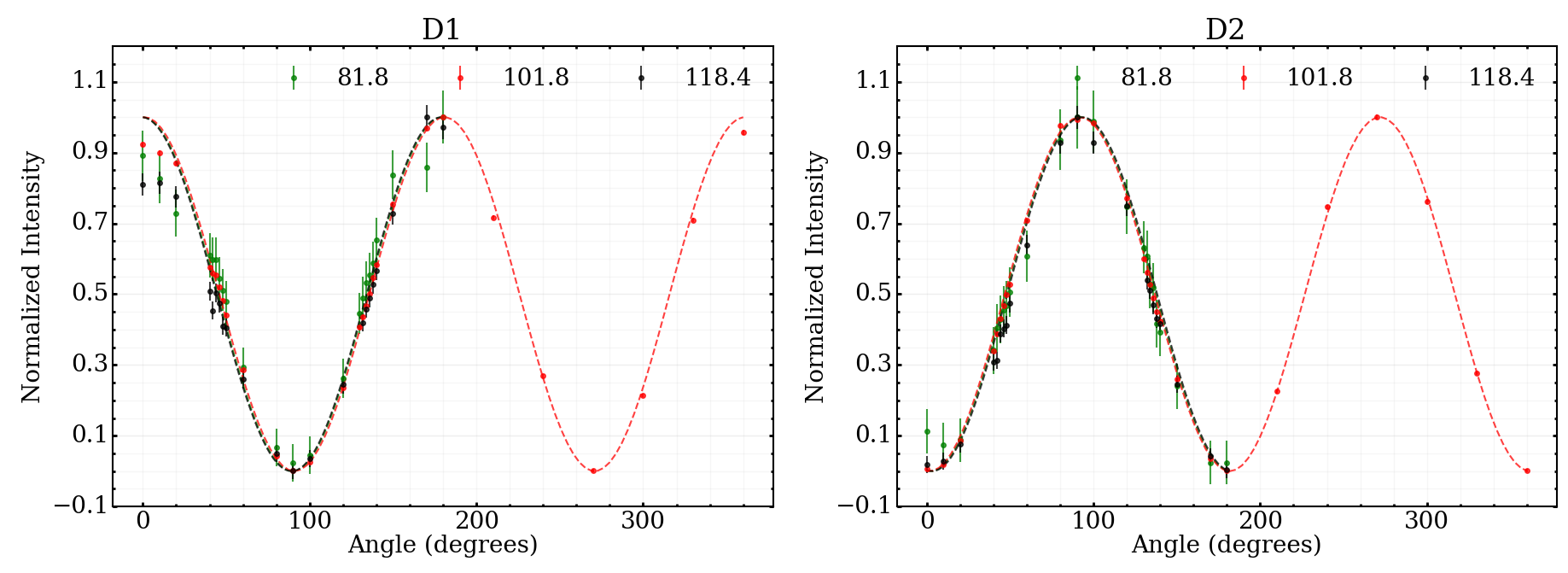}
\caption{Variation of the normalized intensity for Detector-1 (left) and Detector-2 (right) with polarizer angle. The data was recorded at 81.8 MHz (green), 101.8 MHz (red) and 118.4 MHz (black). The markers are the laboratory data with errorbars and the dashed line is the best-fit. At 81.8 MHz and 118.4 MHz, data is shown for angles $0^\mathrm{o}-180^\mathrm{o}$ and for 101.8 MHz data is shown for angles $0^\mathrm{o}-360^\mathrm{o}$.}
\label{fig:pol_cal}
\end{figure}

AOTFs are excellent polarizers having polarimetric efficiencies $\beta>0.99$ \citep{2022JATIS...8d4007J}. To demonstrate the polarimetric capability, a broadband source followed by a polarizer was kept at the input of the instrument. The rotation modulation of the linearly polarized light in two orthogonally polarized channels of SHAPE was measured for three frequencies (81.8, 101.8 and 118.4 MHz corresponding to 1680, 1363 and 1182 nm respectively) in the operating range and the same is shown in Figure \ref{fig:pol_cal}. Finer measurements were carried out in the $0^\mathrm{o}-180^\mathrm{o}$ and coarser measurements were carried out in the $180^\mathrm{o}-360^\mathrm{o}$. The polarimetric efficiency derived from the fits using the modulation depth is $>0.99$ for all three wavelengths across the operating range and is in good agreement with the values reported in  \cite{2022JATIS...8d4007J}. The polarimetric property is largely constant with wavelength.

\subsubsection{Field Calibration}
\begin{figure}[h!]
\centering
\includegraphics[scale=0.53]{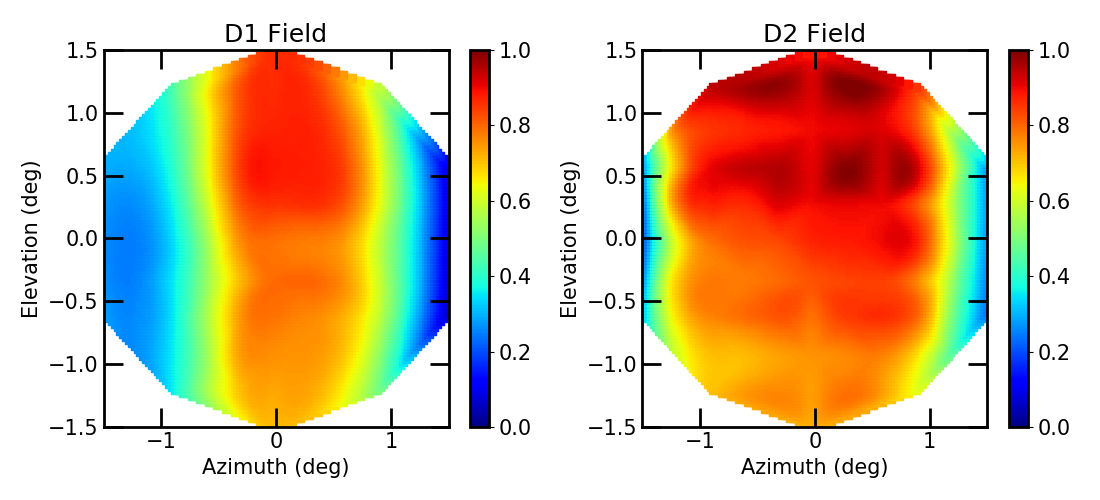}
\caption{The interpolated field variation for both detectors at 1400 nm. The efficiency of Detector-2 is higher than Detector-1 and the field for Detector-2 is relatively more uniform than Detector-1.}
\label{fig:field_cal}
\end{figure}

The FOV of the instrument is $\sim 2.6^\mathrm{o}$ to cover the entire disc of the Earth. The flatness of the FOV in the two channels was measured to obtain the correction factors for a large range of azimuth and elevation angles ($\pm 1.5^\mathrm{o}$ in steps of $0.25^\mathrm{o}$). To carry out the field calibration, the instrument was mounted on a rotation stage to vary the input angles of the source. Broadband un-polarized light was used as input and a complete frequency scan was recorded for each combination of azimuth and elevation angles. The field measured in the lab at one wavelength has been presented in \cite{2024arXiv241207416N} and the details about the setup and methodology to estimate the field response will be presented the onboard calibration paper (Jaiswal et al., under review). The ground measurements were carried out in steps of $0.25^\mathrm{o}$ and the full field is then interpolated using 2-dimensional spline interpolation, which is shown in Figure \ref{fig:field_cal} where the obtained values were normalized for the maximum in Detector-2. This also shows that the beam falling on Detector-2 has better efficiency as compared to Detector-1. However, for both detectors the efficiency falls sharply towards the edge in the azimuthal direction (larger gradient for Detector-1) while is relatively constant in the elevation direction. The field response of the AOTF-based spectro-polarimeter exhibits a degree of non-linearity which can be attributed to the angular dependence of the acousto-optic interaction and minor misalignment between the various optical elements. The field response may change due to vibrations experienced during launch. Efforts to estimate the on-board field response are subject to significant limitations and are discussed in detail in Jaiswal et al. (under review).

\subsection{Spacecraft Level Tests}
\begin{figure}[h!]
\centering
\includegraphics[scale=0.45]{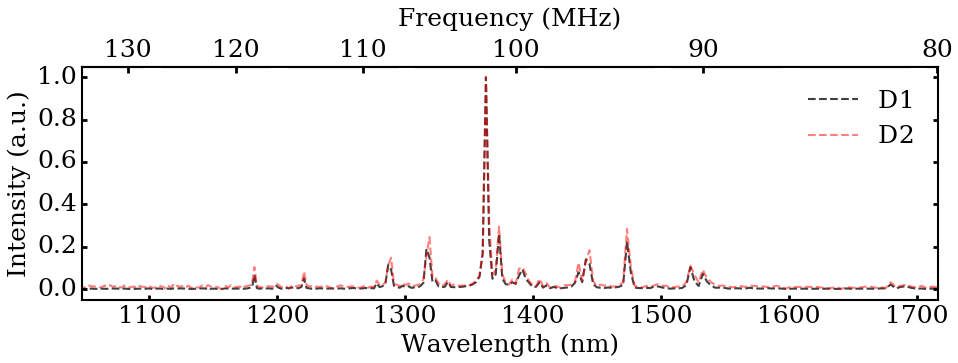}
\caption{Krypton lamp spectrum recorded for both detectors after the instrument was integrated with the spacecraft. The bottom and top axes of the plot are wavelength and frequency input to the AOTF, respectively. The emission lines detected in the spectrum were used for the full functional verification of the instrument.}
\label{fig:ait_spec}
\end{figure}

After the Subsystem level tests were completed, the instrument was integrated with the satellite and various spacecraft level tests were carried out. These included tests to check the performance of the instrument when it was mounted on the spacecraft and to understand the influence of the instrument on other subsystems mounted on the spacecraft and vice versa. After these tests, dynamic testing of the spacecraft was carried out. A test was carried out to check the performance of the payload after the dynamic tests of spacecraft. All these tests were carried out using a Krypton Lamp as the source and the spectrum was recorded in both modes of operation with varying integration time and input RF power. The spectrum in the Sum Mode of operation is shown in Figure \ref{fig:ait_spec}. These tests demonstrated the end-to-end performance verification of the instrument after integration with the spacecraft.

\section{In-flight Operation and Observation}
There are two orbital phases of the Ch3-PM -- the Lunar orbiting phase till $\sim$ mid-October 2023, and the Earth orbiting phase after that. In the Lunar orbiting phase, the Ch3-PM orbited the Moon in an orbit of $\sim$~$148$~km $\times$ $153$~km with an orbital period of $\sim$~$2$~hr. The SHAPE payload on the Ch3-PM was commissioned on $20$ August 2023, after the separation of the LM from the PM. The SHAPE operation involved maneuvering the Ch3-PM to get its +Roll axis, along which the SHAPE's optical view axis is aligned, to point towards Earth to carry out the spectro-polarimetric observations (see Figure 5 in \cite{2022JATIS...8d4007J} for the observation geometry). In October 2023, the orbit of the propulsion module was raised to get it back into a target Earth-bound orbit with an altitude range of approximately 180,000 km $\times$ 380,000 km. Given this highly elliptical orbit, the orbital period in this Earth-bound phase is $\sim$~12 days. Only near the apogee of the orbit, the angle subtended by Earth is suitable for observations with SHAPE. The spacecraft's perigee and apogee vary throughout its trajectory, with the minimum predicted perigee altitude reaching approximately 115,000 km. Along with the science observations of Earth, SHAPE also carried out several calibration observations of the Moon from the HEO (Jaiswal et al., in prep). In addition to these observations, SHAPE also carried out two observations of the Sun during solar eclipse ingress, but the detectors were saturated in both observations. The PM entered the Moon’s Sphere of Influence on 4 November 2025, and subsequently executed two lunar flybys on 6 and 14 November 2025. As a result of these maneuvers, the PM's HEO around Earth was modified to 409,000 km $\times$ 727,000 km. Post this, SHAPE has continued to observe Earth. A summary of the SHAPE observations is provided in Table \ref{tab:obs_sum}. As the HEO evolves over time and is not maintained, Table \ref{tab:obs_sum} lists the initial orbital sizes.

\begin{table}[h!]
\begin{center}
\caption{Summary of SHAPE observations till the 18 August 2026.}\label{tab:obs_sum}
\begin{tabular}{|c|c|c|c|}
\hline
Orbit               									& Object     	& Number of Observations  & Observation Type 	\\ \hline
LO (153 km $\times$ 163 km)														& Earth      	  & 72                     & Science          	\\ \hline
\multirow{2}{*}{HEO (180,000 km $\times$ 380,000 km)} 	& Moon      	& 19                      & Calibration      	\\ \cline{2-4} 
                    									& Earth      	& 48                      & Science         	 	\\ \hline
HEO (409,000 km $\times$ 727,000 km)					& Earth      	& 7						  & Science         	 	\\ \hline                    	
\end{tabular}
\end{center}
\end{table}

During each observation, first the payload is thermally stabilized for a few minutes after switching on, then the science observations are carried out once the payload axis is pointed in the direction of the object of interest (Earth/Moon), followed by the data playback via the telemetry channel. Each science operation involves the acquisition of multiple Earth spectra. During each science operation the PM drifts from the initial pointing which is coarsely corrected for by firing thrusters. Due to this the object moves in the FOV. The raw data is downloaded at the Indian Space Science Data Centre (ISSDC), ISRO and processed to generate the Level-0 product. The Level-0 product is transferred to the payload operation centre for generating higher level science ready products. Each science observation also consists of pre- and post-background measurements, which will be used to subtract the background from the science observations while generating the science ready products. The operation plan for SHAPE will be discussed in detail in a subsequent paper. During the science observations, the Earth-Sun angle with respect to SHAPE's optical axis is maintained to be greater than 5$^\mathrm{o}$ to avoid the saturation of detectors. The Earth observations are carried out for various phase angles ranging from $\sim0^\mathrm{o} - 180^\mathrm{o}$. As the signal from Earth decreases as the phase increases \citep{2020A&A...640A.121G}, the observation would be carried out using higher integration times for larger phase angles. The spacecraft also provides the object (Moon/Earth) pointing geometry, including the angular coordinates required to locate the object within the instrument FOV, as well as the sub-PM latitude and longitude on the object's surface. These can be used to obtain the specific view of the object observed by the instrument. Both spectral data and the associated angles are time-tagged in the Coordinated Universal Time (UTC). A detailed description of these data products and their processing will be presented in a forthcoming pipeline paper (Ravishankar et al., in preparation). This information is required to understand the regions of Earth viewed by SHAPE and use simultaneous cloud properties from the Moderate Resolution Imaging Spectroradiometer (MODIS; \citealp{2003ITGRS..41..442K}) to model the SHAPE spectro-polarimetric data. 

\begin{figure}[h!]
\centering
(a)\includegraphics[scale=0.33]{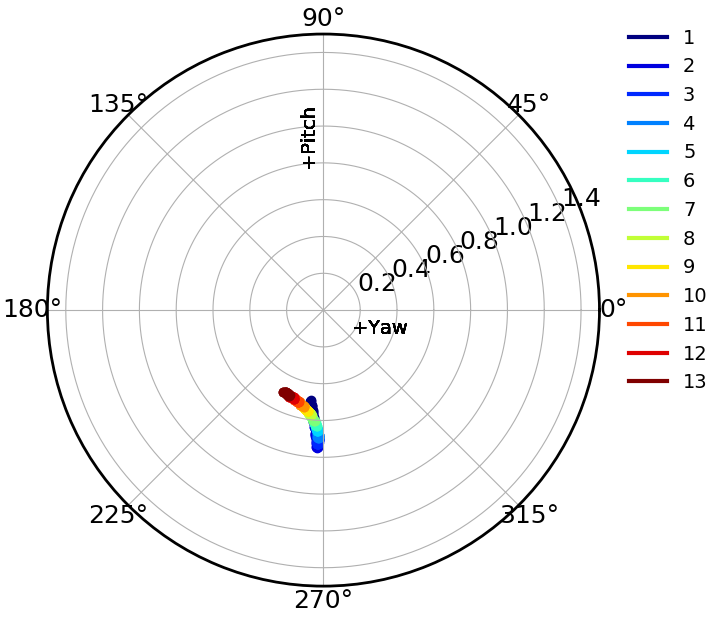}\hspace{8mm}
(b)\includegraphics[scale=0.37]{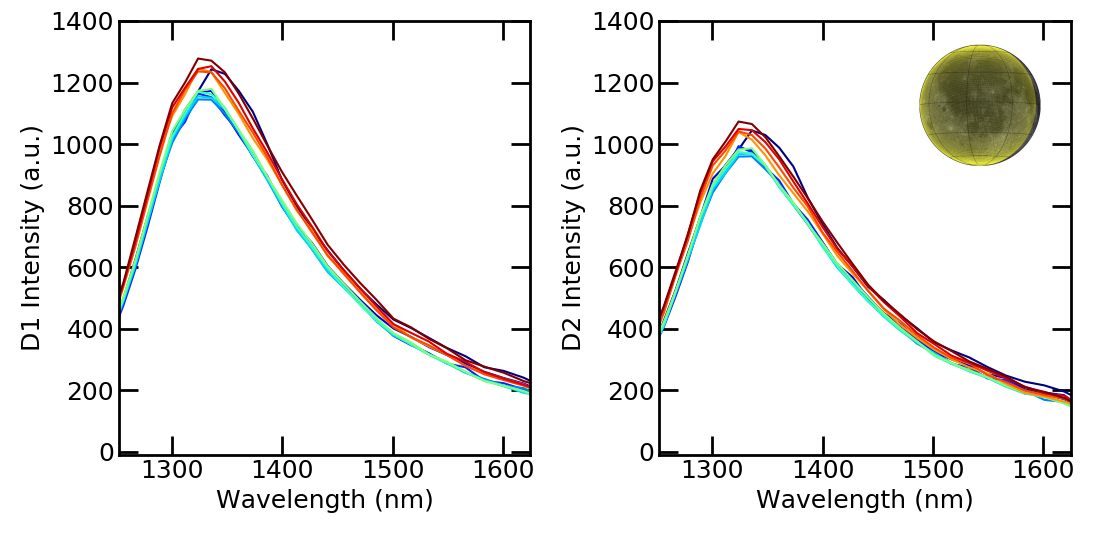}\vspace{7mm}
(c)\includegraphics[scale=0.31]{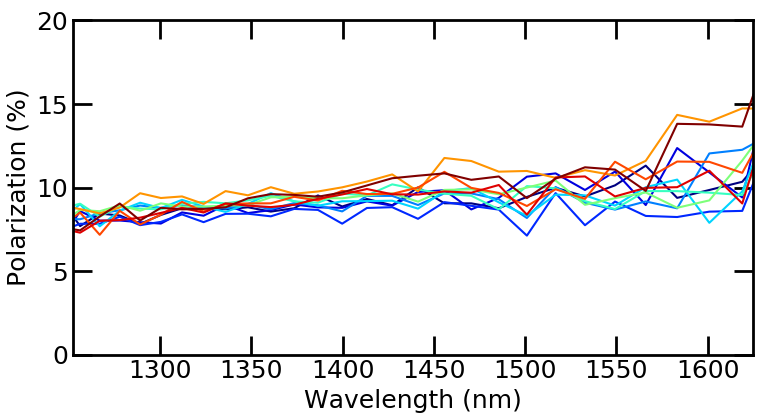}\hspace{7mm}
(d)\includegraphics[scale=0.305]{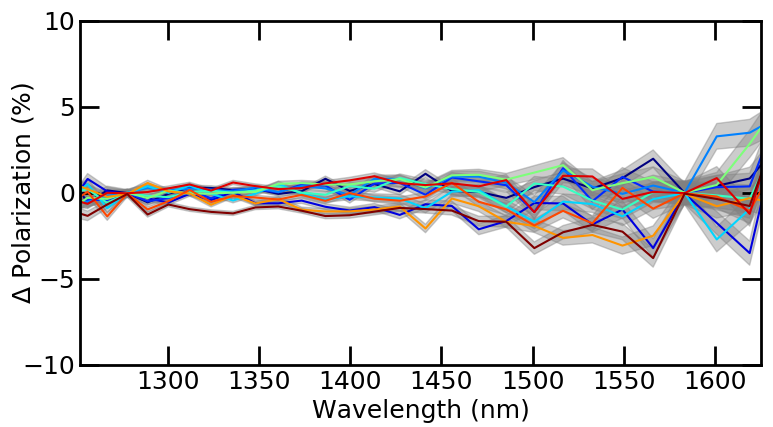}
\caption{(a) Movement of Moon within the FOV for Moon observation from HEO on 31 December 2023. The colours corresponding to each spectrum (spectrum number $1-13$) indicated here are followed across all plots in this figure. For spectra 8 and 9, only the location in the FOV is shown, as the spectra were saturated and are therefore omitted from the subsequent plots. (b) SHAPE spectrum recorded for Detector-1 (left) and Detector-2 (right). The inset shows the Moon view for this observation and the yellow shaded region shows the sun-illuminated regions of the Moon. (c) Absolute polarisation measurements indicating the effect of instrument induced polarisation. (d) $\triangle$ polarisation measurements indicating the expected flat polarisation. The grey shaded regions indicate the errors on the $\triangle$ polarisation which represent the propagated statistical uncertainty ($1\sigma$) on the measured $\triangle$ polarization, obtained from the photon-counting statistics and detector noise in the two orthogonally polarized channels. As the instrument response decreases towards longer wavelengths (especially $>1600$ nm), the detected signal level decreases, resulting in reduced SNR and consequently larger statistical uncertainty.}
\label{fig:sci_spec_moon}
\end{figure}

There have been several Moon observations by SHAPE from the HEO. Figure \ref{fig:sci_spec_moon}(a) and (b) show the movement of Moon in the FOV and the SHAPE spectra recorded for both the detectors corresponding to a lunar observation carried out on 31 December 2023, with the inset showing the view of the Moon seen by the payload. The SHAPE spectra from lunar observations consist of reflected solar continuum shaped by lunar surface reflectance and the instrument response, which clearly shows a decline in the efficiency towards both ends of the spectral range. With the help of this observation, we show the systematic variation observed in estimating the polarisation in Figure \ref{fig:sci_spec_moon}(c) and then also demonstrate the capability of SHAPE to study the band polarisation (or $\triangle$ polarisation). $\triangle$ polarisation is defined as the wavelength-dependent deviation of the measured degree of linear polarisation from the underlying linear continuum polarisation trend. As mentioned previously, from the lab measurements we see that the SHAPE field has a non-uniform response across the FOV which results in the instrument's polarimetric response to vary across the FOV. This, combined with the random drift of the spacecraft (upto $\sim 1^\circ$) produces polarisation which is affected by instrumental polarisation and has variation across different regions of the field. As the Moon as well as Earth are extended sources, in addition to the source brightness and integration time, the SNR depends on its apparent angular size and its location within the FOV. Thus, the polarimetric precision for these observations varies from one observation to another depending on the observing conditions. Moon observations, however, have also demonstrated the linear trends in the spectral polarisation, which, if removed can be useful to study the spectral-band polarisation of Earth. The Earth, having atmosphere and oceans can have band-polarisation \citep{2022A&A...664A.172T}. The Moon, however, is devoid of atmosphere and oceans and is expected to have a flat spectral polarisation \citep{1971A&A....10...29D}. In Figure \ref{fig:sci_spec_moon}(d), we demonstrate the estimation of flat spectral-band polarisation. The details of instrument characteristics and on-board polarimetric performance are comprehensively discussed in a separate research article (Jaiswal et al., under review).

\begin{figure}[h!]
\centering
\includegraphics[scale=0.50]{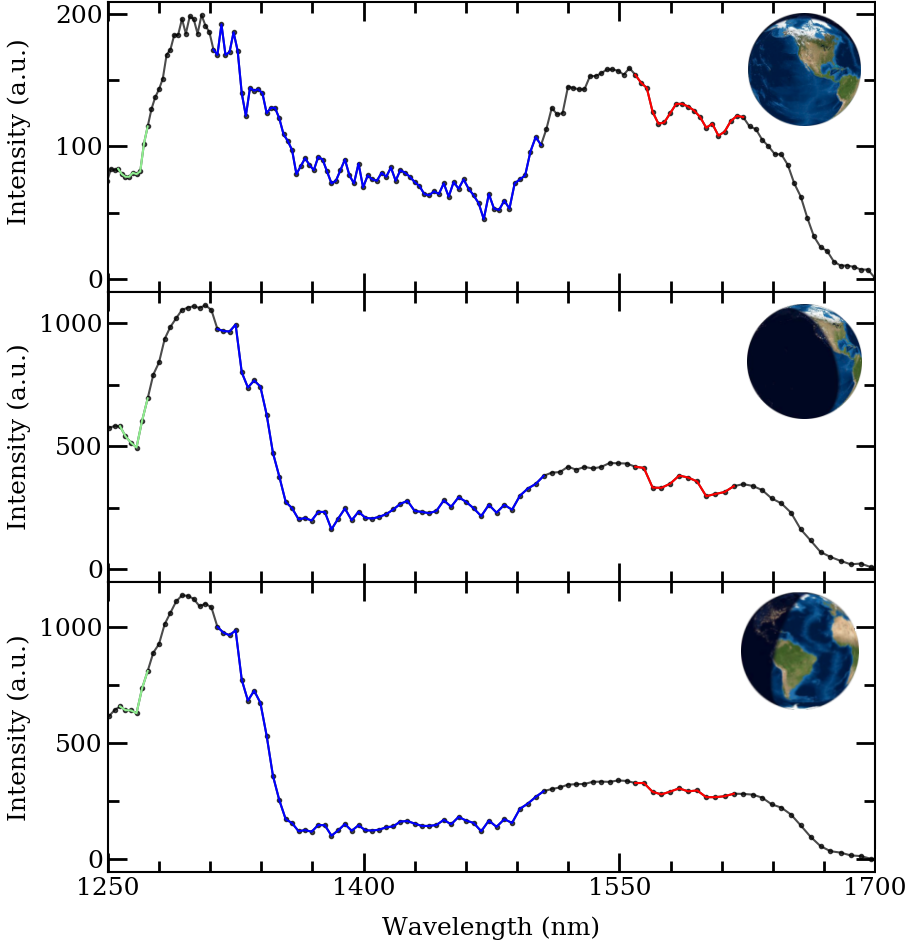}
\caption{SHAPE spectra recorded for Earth observations from LO on 03 October 2023 (top), from HEO (180,000 km $\times$ 380,000 km) on 17 April 2025 (middle) and from HEO (409,000 km $\times$ 727,000 km) on 10 February 2026 (bottom). The Earth view at the time of observation is shown in the insets of the respective spectra. The green, blue and red highlighted regions show the O$_2$, H$_2$O and CO$_2$ bands.}
\label{fig:sci_spec_earth}
\end{figure}

Figure \ref{fig:sci_spec_earth} shows the SHAPE spectra recorded for Earth in various orbits, with the Earth view for each observation shown as an inset. The instrument's spectral resolution, wavelength coverage and sensitivity are designed to resolve and detect characteristic absorption bands of atmospheric species in the Earth's atmosphere. The figure shows the detection of well-known atmospheric absorption bands of H$_2$O, O$_2$ and CO$_2$ in SHAPE observations, confirming that the instrument provides the required wavelength coverage, spectral resolution and sensitivity needed to resolve and measure these spectral features. The successful identification of these molecular bands in both horizontal and vertical polarisations demonstrates that the instrument can detect wavelength-dependent spectral-polarimetric signatures across various orbits. These observations validate the end-to-end performance (consistent with the SHAPE Capabilities listed in Table \ref{table:scireq}) of the optical system, detector, calibration procedures and data reduction pipeline. The per-pixel noise and dark signal were found to be consistent with the on-ground measurements. Using the continuum region free from significant atmospheric absorption features in Figure \ref{fig:sci_spec_earth}, the achieved SNR was estimated to be greater than 100. SHAPE observations of disc-integrated Earth confirm the detection of these biosignature gases at various phase angles across all three orbits (see Table \ref{tab:obs_sum}). Consistent with previous measurements, we detect a broad 1.4 \textmu m H$_2$O absorption band, the prominent O$_2$ band centered at 1.27 \textmu m, and two CO$_2$ absorption features near 1.57 \textmu m and 1.61 \textmu m. The relative depths and shapes of these absorption features vary with viewing geometry, reflecting changes in cloud cover, surface albedo and atmospheric path length as a function of phase angle. The detailed observational trends and their physical interpretations are currently under investigation and will be presented in subsequent dedicated science publications.

\section{Summary and Conclusion}
The SHAPE instrument onboard the Chandrayaan-3 PM has been successfully designed, developed, and operated to conduct spectro-polarimetric observations of Earth from both lunar and highly elliptical Earth-bound orbits. This paper presented an overview of the instrument's design, its integration and testing, and initial in-orbit performance. The results demonstrate that SHAPE is capable of capturing disc-integrated polarimetric signatures of Earth over a range of phase angles, validating its suitability as a technology pathfinder for future exoplanet characterization missions. Till date, SHAPE has been operated over 146 times. SHAPE has conducted interesting and scientifically valuable observations of the Earth, wherein the instrument successfully detected the O$_2$, H$_2$O, and CO$_2$ absorption bands, in addition to performing detailed measurements of the polarisation characteristics within the H$_2$O band. 
Detailed results pertaining to onboard calibration procedures, the end-to-end data processing pipeline, and the first scientific analyses will be discussed in forthcoming publications (Jaiswal et al., under review and Ravishankar et al., in preparation). The SHAPE instrument continues to operate nominally and will acquire scientific data whenever suitable observational geometry and operational conditions are achieved.

\section*{Acknowledgments}
Authors thank the reviewers for their comments and constructive suggestions. Authors thank the Group Head, Space Astronomy Group; Deputy Director, Payload Data Management \& Space Astronomy Area; Associate Director and Director of U. R. Rao Satellite Centre for encouragement and continuous support to carry out this work. Authors also thank Group Director, Communication Systems Group; Group Director, Thermal Systems Group; Group Director, Systems Integration Group; and Director, LEOS for constant support towards realization of the payload. The Indian Space Research Organization (ISRO) funded, managed and facilitated the overall project. 

\section*{Data Availability}
The observational data used will be made available at \url{https://pradan.issdc.gov.in/ch3} by the instrument team and the Indian Space Science Data Center (ISSDC).

\bibliographystyle{jasr-model5-names}
\biboptions{authoryear}
\bibliography{refs}

\end{document}